\documentclass{article}

\usepackage{amsmath,amsfonts,amssymb,mathtools,amsthm}
\usepackage{stmaryrd}
\usepackage{tikz}
\usepackage{pgfplots}
\pgfplotsset{width=16cm}
\usepackage{geometry} 
 
\newtheorem{assumption}{Assumption}{\bfseries}{\itshape}
{\bfseries}{\itshape}
\newtheorem{proposition}{Proposition}

\newtheorem{heuristic}{Heuristic}
\newtheorem{lemma}{Lemma}

\newtheorem{modeling}{Modeling}

\newtheorem{definition}{Definition}

\theoremstyle{remark}
\newtheorem{remark}{Remark}

\newcommand{\mat}[1]{\ensuremath{\boldsymbol{#1}}}
\newcommand{\card}[1]{\lvert #1 \rvert}

\newcommand{\Am}{\mat{A}}

\newcommand{\uv}{\mat{u}}
\newcommand{\vv}{\mat{v}}

\newcommand{\xv}{{\mat{x}}}
\newcommand{\yv}{{\mat{y}}}
\newcommand{\zerov}{{\mat{0}}}
\newcommand{\zv}{{\mat{z}}}

\newcommand{\Mac}{\text{Mac}}

\newcommand{\F}{\mathbb{F}}
\newcommand{\N}{\mathbb{N}}
\newcommand{\R}{\mathbb{R}}
\DeclareMathOperator{\HS}{HS}
\DeclareMathOperator{\HF}{HF}
\DeclareMathOperator{\LM}{LM}

\DeclareMathOperator{\cI}{\mathcal{I}}
\DeclareMathOperator{\cQ}{\mathcal{Q}}
\DeclareMathOperator{\cR}{\mathcal{R}}
\DeclareMathOperator{\cAMR}{\mathcal{AMR}}
\DeclareMathOperator{\supp}{supp}
\DeclareMathOperator{\ev}{ev}
\newcommand{\Iintv}[2]{\llbracket #1 , #2 \rrbracket}
\newcommand{\OneN}[1]{\llbracket #1 \rrbracket}
\newcommand{\floor}[1]{\left\lfloor #1 \right\rfloor}

\newcommand{\eqdef}{\stackrel{\text{def}}{=}}

\title{The regular multivariate quadratic problem}
\author{Antoine Joux, Rocco Mora}
\date{CISPA Helmholtz Center for Information Security - Germany}

\begin{document}

\maketitle

\begin{abstract}
In this work, we introduce a novel variant of the multivariate quadratic problem, which is at the core of one of the most promising post-quantum alternatives: multivariate cryptography. In this variant, the solution of a given multivariate quadratic system must also be regular, i.e. if it is split into multiple blocks of consecutive entries with the same fixed length, then each block has only one nonzero entry. We prove the NP-completeness of this variant and show similarities and differences with other computational problems used in cryptography. Then we analyze its hardness by reviewing the most common solvers for polynomial systems over finite fields, derive asymptotic formulas for the corresponding complexities and compare the different approaches.
\end{abstract}
\section{Introduction}

\textbf{The MQ problem and its use in cryptography.} The problem of solving a multivariate quadratic system over a finite field takes the name of Multivariate Quadratic (MQ) problem and is known to be NP-complete. It also seems to be hard on average for a wide range of parameters, but despite its evident difficulty, numerous algorithms have been developed to solve it. 
The MQ problem is fundamental in cryptography, especially in the post-quantum realm. It serves as the foundation for designing multivariate public key cryptosystems, which form one of the most promising quantum-resistant alternatives. In particular, it has emerged over the years as being remarkably well-suited to the construction of digital signatures. It is even more pervasive in algebraic cryptanalysis, which has a crucial impact on many different families of primitives. Indeed, algebraic attacks aim at solving a polynomial system that models a cryptographic problem. 

\textbf{A new variant of the MQ problem.} We present here a new variant of the MQ problem. Whilst the polynomial equations are sampled at random and do not present any particular structure, the solution is required to have a regular shape. Concretely, if the solution is a vector of $n=lw$ variables split into $w$ blocks each of length $l$, then it must have exactly one nonzero entry for each block. The NP-completeness of the multivariate quadratic problem is preserved when adding the regular condition to the solution.

The same regular constraint was originally introduced in the context of code-based cryptography in \cite{AFS05}. Indeed, \cite{AFS05} proposed a family of cryptographic hash functions based on the provably secure regular syndrome decoding (RSD) problem. More recently, the same problem received new attention thanks to its introduction in secure computation to design MPC-protocols \cite{HOSS22}, pseudorandom correlation generators \cite{BCGI18} and digital signatures from MPC-in-the-head \cite{CCJ23, BCCGJ24} or VOLE-in-the-head \cite{CLYYYZ24}.
These works suggest that also the RMQ problem could be used for similar constructions. On the other hand, there have been attempts at improving the cryptanalysis, either with combinatorial techniques \cite{CCJ23, ES24} or with algebraic modelings \cite{BO23}.
In continuity with its code-based predecessor, the new variant has been named the Regular Multivariate Quadratic (RMQ) problem.

\textbf{Our contribution.} 
After introducing the computational problem and overviewing its main features, we mainly focus on the study of the asymptotic complexity of the average RMQ instance in the setting of the uniqueness bound. We analyzed the most important families of techniques for solving boolean systems and adapted them to take advantage of the regularity of the solution. On one side, we explore the algebraic algorithms related to Gr\"obner basis computation \cite{CKPS00, BFSS13} by modeling the regular constraint as additional equations to be satisfied, reviewing different ways of hybridization and estimating the solving degree, in a similar fashion to what has been done for the code-based counterpart in \cite{BO23}. On the other hand, we investigate how the research line of probabilistic polynomial methods initiated with \cite{LPTWY17} applies and can be optimized to our context. We also show how an RMQ instance can be transformed into a standard MQ one with fewer variables and analyze the complexity of solving the new system. Finally, we provide comparisons among all the approaches examined. Despite the improvements presented, we conclude that the problem remains difficult in practice in the average case for a large range of parameters and therefore it is potentially suitable for cryptographic purposes.

\section{Notation and preliminaries}
\subsection{Notation}

\textbf{Sets and intervals.} The closed interval of real numbers between $a$ and $b$ is denoted by $[a,b]$, while the corresponding interval of integers by $\Iintv{a}{b}$ and if $a=1$ we simply write $\OneN{b}$. The cardinality of a set $S$ is indicated with $\card{S}$. 

\textbf{Vectors, matrices and polynomials over finite fields.} Given a prime power $q$, we denote by $\F_q$ the finite field of size $q$. The length-$n$ linear space over $\F_q$ is $\F_q^n$ and the vector space of matrices of size $n\times m$ over $\F_q$ is $\F_q^{n\times m}$. Vectors (respectively matrices) are written with lowercase (respectively uppercase) bold letters, while their entries are not bold, for instance $\vv=(v_1,\dots,v_n)\in \F_q^n$ and $\Am=(a_{i,j})\in \F_q^{n\times m}$. We denote by $\binom{\OneN{n}}{w}$ (respectively $\binom{\OneN{n}}{\downarrow w}$) the sets of vectors in $\F_2^n$ of weight exactly $w$ (respectively at most $w$). It holds that $\card{\binom{\OneN{n}}{\downarrow w}}=\sum_{i \in \Iintv{0}{w}} \binom{n}{i}$ and we denote such sum with $\binom{n}{\downarrow w}$.

Finally, the polynomial ring in a vector of $n$ variables $\xv=(x_1,\dots,x_n)$ over $\F_q$ is written as $R=\F_q[\xv]$ and its polynomials as $P(\xv)$. We use calligraphic upper-case letters for polynomial ideals and angle brackets to denote ideals generated by sequences or sets of polynomials, e.g. $\cI=\langle S\rangle =\langle f_1,\dots,f_m\rangle$.

\textbf{Regular vectors.} Let $n=wl$, for two positive integers $w$ and $l$. An \textbf{$l$-regular vector} $\vv\in\F_q^n$ is a vector such that 
\[
\forall i \in \Iintv{0}{w-1},\quad \card{\{v_{il+j} \mid j \in \Iintv{1}{l}, v_{il+j}\ne 0}\}=1,
\]
i.e. such that in each of the $w$ windows of entries of length $l$, \textit{exactly one} coordinate is non-zero. The Hamming weight of an $l$-regular vector is thus $w=n/l$. 
We denote by $\cR_l(\F_2^n)$ the set of such vectors and more in general by $\cR_l(S)$ the set of regular vectors within the set $S$. 
We also say that a vector $\vv\in\F_q^n$ is \textbf{$l$-at most regular} if
\[
\forall i \in \Iintv{0}{w-1},\quad \card{\{v_{il+j} \mid j \in \Iintv{1}{l}, v_{il+j}\ne 0}\le 1,
\]
i.e. such that in each of the $w$ windows of entries of length $l$, \textit{at most one} coordinate is non-zero. The Hamming weight of an $l$-regular vector is thus $\le w=n/l$.  The set of all such vectors is indicated with $\cAMR_l(\F_2^n)$ and more in general the set of regular vectors within the set $S$ is $\cR_l(S)$. We clearly have $\cR_l(\F_2^n)\subset \cAMR_l(\F_2^n)$ and $\card{\cR_l(\F_2^n)}=l^{n/l}$ and $\card{\cAMR_l(\F_2^j)}=(l+1)^{n/l}$. From now on, $l$ will always denote the length of a window, so we simply say ``(at most) regular'', simplify the notation with $\cR$ and $\cAMR$, and assume that $l$ divides the vector length $n$.

For the sake of clarity, we also denote the coordinates of a vector of length $n=lw$, by using two indexes; one for the window (starting from 0) and one for the position within the window (starting from 1). In other words, given a vector $\vv\in \F_q^{lw}$, we use the notation
\[
v_{i,j}\eqdef v_{il+j},\qquad \forall  i \in \Iintv{0}{w-1}, j \in \Iintv{1}{l}.
\]
This notation is applied to vectors of variables as well. We also use the term ``block'' as a synonym for ``window''.

\textbf{Complexity and asymptotic.} Given a function $f$, we use the standard big O notation $\mathcal{O}(f(n))$ to provide the cost of an algorithm or one of its steps, meaning that the cost is upper bounded by $c f(n)$, for a positive real constant $c$ and for any $n\ge n_0$, where $n_0$ is a fixed integer. The metric we adopt uses the number of operations over the target finite field $\F_q$ as unit. For exponential costs, we also make use of the notation $\mathcal{O}^*(f(n))$ to denote the asymptotic behavior, up to polynomial factors in $n$, i.e. the cost is upper bounded by $P(n) f(n)$ for a fixed polynomial $P$ and for any $n\ge n_0$, where $n_0$ is a fixed integer. 

The logarithm $\log(\cdot)$ has to be intended in base 2, while $\ln(\cdot)$ is the natural logarithm. Let $H\colon [0,1]\to [0,1]$ be the binary entropy function defined as
\[
H(0)=H(1)=0
\]
and
\[
H(p)\eqdef -p\log(p)-(1-p)\log(1-p) \qquad \text{otherwise}.
\]
The binary entropy function is increasing in $[0,1/2]$, reaches its maximum $H(1/2)=1$ and it is decreasing in $[1/2,1]$, being null at its extremes. We also define the non-decreasing function
\[
H^*(p)\eqdef \begin{cases}
    H(p), & p \in [0,1/2)\\
    1, & p \in [1/2,1].
\end{cases}
\]
We have that
\[
\binom{n}{w}=\mathcal{O}^*(2^{H(w/n)n}) \qquad \text{and} \qquad \binom{n}{\downarrow w}=\mathcal{O}^*(2^{H^*(w/n)n}).
\]
Finally, the Greek letter $2\le \omega\le 3$ denotes the linear algebra constant. When sparse linear algebra is possible, quadratic-time algorithms allow to replace $w$ with 2.

\subsection{Algebraic cryptanalysis}

Let $R=\F[x_1,\dots,x_n]=\F[\xv]$ be the polynomial ring over the finite field $\F$ in $n$ variables. A polynomial $f\in R$ is said \textit{homogeneous} if all its monomials have the same degree and \textit{affine} otherwise. We denote by $R_d$ the vector space of all homogeneous polynomials of degree $d$, which is generated by all degree-$d$ monomials. We also consider ideals generated by a set (or sequences) of polynomials in $R$. If an ideal $\mathcal{I}\subset R$ admits a set of homogeneous generators we say that the ideal is homogeneous. In this case, we denote with $\mathcal{I}_d \eqdef \mathcal{I} \cap R_d$, i.e. the subspace of $A_d$ consisting of all homogeneous polynomials of degree $d$ in $\mathcal{I}$.

The \textit{Hilbert function} and the \textit{Hilbert series} describe the combinatorial structure of homogeneous ideals (or equivalently the corresponding quotient ring) and represent a powerful tool to study the complexity of solving algebraic systems. 
\begin{definition}[Hilbert function/series]
Let $\mathcal{I}\subseteq R$ be a homogeneous ideal. The \textbf{Hilbert function} is defined as
\begin{align*}
\HF_{R/\mathcal{I}} \colon & \N \to \N \\
& d\mapsto \dim_{\F} R_d/\mathcal{I}_d
\end{align*}
and the \textbf{Hilbert series} as its generating series, i.e.
\[
\HS_{R/\mathcal{I}}(z)\eqdef \sum_{d=0}^{\infty} \HF_{R/\mathcal{I}}(d) z^d.
\]
\end{definition}

Given an ideal $\mathcal{I}$, its variety $\mathcal{V}(\mathcal{I})$ is the set of common zeros for all the elements in $\mathcal{I}$ whose coordinates lie in the algebraic closure of $\F$. However, for cryptographic purposes, one usually wants to find solutions over the field where the constraints are defined or at least over some finite extension. The constraints that force solutions to lie in the finite field of $q$ elements $\F=\F_q$ are the so-called \textbf{field equations}:
\[
x_i^q-x_i=0 \quad \forall i \in \Iintv{1}{n}.
\]
When an ideal $\cI$ contains the field equations, then it is always \textit{zero-dimensional}, i.e. the quotient $R/\cI$ is a finite-dimensional vector space. In this case, the Hilbert series is simply a polynomial, that is there exists a positive integer $d$ such that $\cI_d=R_d$. We call \textbf{degree of regularity} $d_{reg}$ the smallest integer $d$ that satisfies the equality. 

In the most general case, the Hilbert series of an ideal is difficult to compute and the cost of this task is essentially the same as computing a Gr\"obner basis of the associated system.  On the other hand, the Hilbert series can be predicted for systems of equations that satisfy a certain notion of regularity.
\begin{definition}[Regular system]
Let $m\le n$. The homogeneous system $S=\{f_1,\dots,f_m\}\subseteq R=\F[x_1,\dots,x_n]$ is \textbf{regular} if 
\[
\forall i \in \Iintv{1}{m},\quad f_i \text{ is not a zero divisor in } R/\langle f_1,\dots,f_{i-1}\rangle.
\]
\end{definition} 
\begin{remark}
    The notion of \textit{regularity} of an algebraic system should not be confused with that of \textit{regular} constraint of a solution, with which it only shares the term.
\end{remark}
\begin{proposition}
Let $S=\{f_1,\dots,f_m\}\subseteq R$ be a regular system with $\deg(f_i)=d_i$ for all $i \in \Iintv{1}{m}$. Then
\[
\HS_{R/\langle S\rangle} (z)=\frac{\prod_{i=1}^m (1-z^{d_i})}{(1-z)^n}.
\]
\end{proposition}
The definition of regularity does not cover the overdetermined case, i.e. when $m>n$. The notion has first been extended to this setting in \cite{B04}.
\begin{definition}[Semi-regular sequence]  \label{def: semireg}
Let $S=\{f_1,\dots,f_m\}\subseteq R$ be a homogeneous sequence such that $\cI=\langle S\rangle$ is zero-dimensional and has degree of regularity $d_{reg}$. The sequence $S$ is called \textbf{semi-regular} if $\cI\neq R$ and if for $i \in \Iintv{1}{m}$, $g_i f_i=0$ in $R/\langle f_1,\dots,f_{i-1}\rangle$ with $\deg(g_i f_i)< d_{reg}$ implies $g_i=0$ in $R/\langle f_1\dots,f_{i-1}\rangle$.
\end{definition}
A special definition is given for boolean systems, by taking into account the Frobenius morphism. To this purpose, we define $R'=\F_2[\xv]/\langle x_1^2,\dots,x_n^2\rangle$.
\begin{definition}[Semi-regular sequence over $\F_2$] \label{def: semiregF2}
Let $S=\{f_1,\dots,f_m\}\subseteq R'$ be a homogeneous sequence such that $\cI=\langle S\rangle$ has degree of regularity $d_{reg}$. The sequence $S$ is called \textbf{semi-regular over $\F_2$} if $\cI\neq R'$ and if for $i \in \Iintv{1}{m}$, $g_i f_i=0$ in $R'/\langle f_1,\dots,f_{i-1}\rangle$ with $\deg(g_i f_i)< d_{reg}$ implies $g_i=0$ in $R'/\langle f_1\dots,f_{i-1}\rangle$.
\end{definition}

The Hilbert series is determined for semi-regular sequences as well:
\begin{proposition}
Let $S=\{f_1,\dots,f_m\}$ be a homogeneous semi-regular sequence with $\deg(f_i)=d_i$ for all $i \in \Iintv{1}{m}$. Then
\[
\HS_{R/\langle S\rangle} (z)=\left[\frac{\prod_{i=1}^m (1-z^{d_i})}{(1-z)^n}\right]_+,
\]
where $[\cdot]_+$ denotes the truncation of the series starting from the first non-positive coefficient.

Let $S=\{f_1,\dots,f_m\}$ be a homogeneous semi-regular sequence over $\F_2$ with $\deg(f_i)=d_i$ for all $i \in \Iintv{1}{m}$. Then
\[
\HS_{R'/\langle S\rangle} (z)=\left[\frac{ (1+z)^n}{\prod_{i=1}^m (1+z^{d_i})}\right]_+.
\]

\end{proposition}
We remark that the fact that the Hilbert series is a polynomial (because of the truncation) is in accordance with the ideal being zero-dimensional.

\subsection{Affine systems}
In cryptography, we often deal with affine sequences. In this case, we say that a sequence is semi-regular if $S^{(h)}\eqdef \{f_1^{(h)},\dots,f_m^{(h)}\}$ is, where $f^{(h)}$ denotes the part of highest degree of $f$. It is conjectured (see Fr\"oberg conjecture \cite{F85}) that most systems behave in this way. Informally speaking, if one picks a random polynomial system of $m\le n$ ($m>n$ respectively), then it is regular (semi-regular respectively) except with negligible probability. On the other hand, if a system has some non-random structure, then the results on the Hilbert series may not hold anymore. This is the case for the systems considered in this work, where the \textit{regular constraint} on the solution can be translated in terms of quadratic monomial (and thus non-random) equations. As a workaround, we can split our system into two parts, in the same fashion as done in \cite{BO23}. One of them is considered to be generic, and the other, expressing the regular constraint, is analyzed separately. We thus make use of an assumption that captures the notion of semi-regularity in our context and allows to derive the Hilbert series for the final ideal. 

To solve the affine system, we make use of the linearization technique of the XL algorithm \cite{CKPS00}, which consists in computing the \textit{Macaulay matrix} of the system $S=\{f_1,\dots,f_m\}$ at some sufficiently high degree $d$ such that there are more independent equations than monomials. We recall the definition of the Macaulay matrix for both homogeneous and affine systems.

\begin{definition}[Macaulay Matrix \cite{M94}]
	Let $S=\{f_1,\dots,f_m\}\subset\F[\xv]$ be an algebraic system such that $\deg(f_i)=d_i$ and let $d$ be a positive integer. If $S$ is homogeneous, the \textbf{(homogeneous) Macaulay matrix} $\Mac(S,d)$ of $S$ in degree $d$ is the matrix whose rows are each indexed by a polynomial $m_j f_i$, for any $f_i\in S$ and any monomial $m_j$ of degree $ d-d_i$, and whose columns are indexed by all the monomials of degree $d$. 
 If $S$ is an affine system, then the \textbf{(affine) Macaulay matrix} $\Mac(S,\le d)$ is the matrix whose rows are each indexed by a polynomial $m_j f_i$, for any $f_i\in S$ and any monomial $m_j$ of degree $\le d-d_i$, and whose columns are indexed by all the monomials of degree $\le d$.
\end{definition}

Echelonizing the Macaulay matrix can be interpreted as taking polynomial combinations of elements in $S$ and it is not surprising that it plays an important role in Grobn\"er basis techniques as well, with which the XL algorithm is indeed strictly related. Moreover, we take advantage of the Wiedemann algorithm \cite{W86, C94}, which exploits the matrix sparseness. The cost of block Wiedemann XL algorithm is estimated in many works of this genre to be
\begin{equation} \label{eq: cost_XL}
3 \cdot r_{\mathsf{AVG}} \cdot\#\mathsf{cols}(\Mac(S,\le d))^2,
\end{equation}
where $r_{\mathsf{AVG}}$ is the average row weight of elements of the Macaulay matrix (which coincides with the average number of terms in polynomials of $S$). On the other hand, $\#\mathsf{cols}(\Mac(S,\le d))$ is the number of columns of the Macaulay matrix and it is upper bounded by $\sum_{i=0}^d \binom{n}{i}$ in the binary case (i.e. when reducing by the quadratic field equations), and by $\sum_{i=0}^d \binom{n+i-1}{i}$ for generic fields. From an asymptotic point of view, Equation~\eqref{eq: cost_XL} means that it is possible to replace the linear algebra constant $\omega$ with the value 2.

For affine systems, especially structured ones, the degree of regularity may not always capture the difficulty of finding a solution, exactly because it ignores the part of the system of lower degree. In this context, a more informative object is the witness degree, originally defined in the binary case in \cite{BFSS13} but here generalized to whatever finite field, as done in \cite{BO23}.
\begin{definition}[Witness degree (adapted from \cite{BFSS13})]
    Let $S=\{f_1,\dots,f_m\}\subseteq \F[\xv]$ be an (affine) sequence of polynomials and $\mathcal{I}=\langle S\rangle$ the ideal that it generates. Let $\mathcal{I}_{\le d}$ and $\mathcal{J}_{\le d}$ be the $\F$-vector spaces defined by
    \[
    \mathcal{I}_{\le d} \eqdef \{ f \in \mathcal{I} \mid \deg(f)\le d\}
    \]
    and
    \[
    \mathcal{J}_{\le d} \eqdef \{ f \in \mathcal{I} \mid f=\sum_{i=1}^m  g_i f_i \land \forall i\in \Iintv{1}{m}, \deg(f_i g_i)\le d\}.
    \]
    The \textbf{witness degree} $d_{wit}$ of $S$ is the smallest integer $d_0$ such that $\mathcal{I}_{\le d}=\mathcal{J}_{\le d}$ and $\langle \{\LM(f) \mid f\in \mathcal{I}_{\le d_0}\} \rangle =\LM(\mathcal{I})$, where $\LM(f)$ denotes the leading monomial of $f$ according to some graded order and  $\LM(\mathcal{I})$ is the set of leading monomials of all nonzero element in $\mathcal{I}$.
\end{definition}

Taking the part of highest degree in each polynomial is one way to obtain a homogeneous system from an affine one. Another strategy consists of adding a variable $y$ and, given an affine polynomial $f$, defining the polynomial
\[
f^{(y)}(x_1,\dots,x_n,y)=y^{\deg(f)}f\left(\frac{x_1}{y}, \dots, \frac{x_n}{y}\right).
\]
Then $f^{(y)}$ is a homogeneous polynomial of degree $\deg(f)$ in $n+1$ variables and we call $S^{(y)}=\{f_1^{(y)},\dots,f_m^{(y)}\}$ the \textbf{homogenization} of $S$. Then the degree of regularity of $S^{(y)}$ (which is in general different from that of $S$) provides under some conditions an upper bound on the witness degree. Again, we provide the adaptation of \cite{BO23} to general finite fields of the original results of \cite{BFSS13}, even if our main interest is the field $\F_2$.
\begin{proposition} \label{prop: dwit}
    Let $S=\{f_1,\dots, f_m,x_1^q-x_1, \dots, x_n^q-x_n\}$ be a sequence of polynomials that admits no solutions, and let $\mathcal{I}^{(y)}\eqdef \langle S^{(y)}\rangle$. Then $d_{wit}(S)\le d_{reg}(\mathcal{I}^{(y)})$.
\end{proposition}
The proposition above explains why we estimate the complexity of solving the RMQ problem by deriving the degree of regularity from the Hilbert series of the homogenized system $S^{(y)}$.
The condition in Proposition~\ref{prop: dwit} of having no solutions is in apparent contradiction with the problem. However, it is consistent with the hybrid strategy of the \textsf{BooleanSolve} algorithm \cite{BFSS13} because, for the almost totality of partial guesses, the specialized systems do not admit any solution indeed. The same happens when applying a hybrid approach to an RMQ instance.

\section{The RMQ problem}
We introduce here a new computational problem, called \textbf{regular multivariate quadratic problem}, which is a variant of the classical multivariate problem, but where only regular solutions to the system are admitted. We provide both the decision and search versions of the problem
\begin{definition}[$l$-Regular Multivariate Quadratic ($l$-RMQ) problem]
    Let $l\ge 2, w\ge 1$ and $m\ge 1$ be three integers, $q$ a prime power and $n=wl$. Let $P_1,\dots, P_m$ be $m$ affine quadratic polynomial in $n$ variables $\xv=(x_1,\dots,x_n)$ over the finite field $\F_q$ given as input. Let 
    \begin{equation} \label{eq: system}
    \{P_i(\xv)=0 \mid i \in \Iintv{1}{m}\}
    \end{equation}
    be the associated quadratic system. The $l$-RMQ problem is:
    \begin{itemize}
        \item \textbf{(Decision version)} Determine whether the system \eqref{eq: system} has an $l$-regular solution over $\F_q$
        \item \textbf{(Search version)}
 Find an $l$-regular solution over $\F_q$ of \eqref{eq: system}, if there is any.
    \end{itemize}
\end{definition}
As for the vectors, we will remove the parameter $l$ and simply write ``RMQ problem'' whenever the window length is not relevant. In this article, we mainly focus on the case of boolean system, i.e. when the underlying finite field is $\F_2$ and we sometimes refer to the RMQ problem over $\F_2$ as simply the RMQ problem. Moreover, for what concerns the asymptotic analysis of the problem, we are interested in the setting where the length of the window $l$ is constant and hence the number of windows $w$ grows linearly with $n$, which is the regime where the problem is supposed to be most difficult.

\subsection{On the shape of the quadratic equations for the RMQ problem}
Let 
\[P(\xv)=\sum_{1\le i<j\le n} a_{i,j} x_i x_j+\sum_{1\le i\le n} a_{i} x_i+a\]
be a generic quadratic polynomial. When $P$ is evaluated on a regular (or even at most regular) vector $\vv$, the monomials $x_i x_j$'s, for any pair $(i,j)$ such that $(i-1)\div l =(j-1) \div l$, always vanish, regardless of the coefficients $a_{i,j}$'s, because at least one between $v_i$ and $v_j$ is 0. It is then enough to consider the monomials other than those and redefine, by using the regular-friendly notation,
\begin{equation} \label{eq: P}
    P(\xv)=\sum_{\substack{0\le i_1<i_2\le w-1\\ 1\le j_1,j_2\le l}}a_{i_1,i_2}^{(j_1,j_2)} x_{i_1,j_1} x_{i_2,j_2}+\sum_{\substack{0\le i\le w-1\\ 1\le j\le l}} a_i^{(j)} x_{i,j}+a.
\end{equation}
The memory size of a single quadratic polynomial for the multivariate quadratic problem thus decreases from $\binom{n+1}{2}+1$ to
\[
\binom{n+1}{2}+1-\frac{n}{l}\binom{l}{2}=\frac{n(n-l+2)}{2}+1.
\]
\subsection{On the generation of RMQ instances}
It is possible to construct quadratic systems whose variety contains a given regular vector. Even more, we can construct such a system in a way that all the quadratic equations are indistinguishable from random ones with the shape discussed above. Indeed, let $\vv\in\cR(\F_2^n)$. Let us sample at random over $\F_2$ all the coefficients $a_{i_1,i_2}^{(j_1,j_2)}$'s and $ a_i^{(j)}$'s of $P$ in \eqref{eq: P}. Then we define $a=\sum_{\substack{0\le i_1<i_2\le w-1\\ 1\le j_1,j_2\le l}}a_{i_1,i_2}^{(j_1,j_2)} v_{i_1,j_1} v_{i_2,j_2}+\sum_{\substack{0\le i\le w-1\\ 1\le j\le l}} a_i^{(j)} v_{i,j}$, so that whatever is the value of $\sum_{1\le i<j\le n} a_{i,j} v_i v_j+\sum_{1\le i\le n} a_{i} v_i$, we obtain $P(\vv)=0$. One can easily verify that, if all the coefficients have been sampled at random then, for a fixed $\vv$,
\[
\Pr[a=0]=\Pr\left[ \sum_{1\le i<j\le n} a_{i,j} v_i v_j+\sum_{1\le i\le n} a_{i} v_i=0\right]=1/2.
\]

On the other hand, it is interesting to study for which parameters $n$ and $m$, a regular solution to a \textit{random} quadratic system in $n$ variables and $m=\mu n$ equations is expected. Let us first recall this expectation in the standard setting. The overdetermined regime is when $m>n$. If $\mu>1$, for $n\to \infty$, solutions are not attended, unless with exponentially small probability in $m-n$. On the opposite side, when the system is underdetermined, i.e. when $m<n$, an exponential number of solutions is expected when $n\to \infty$. The intermediate case $\mu=1$ is when the expectation of the number of solutions is 1.

When instead we search for regular solutions, the threshold for the unique solution expected corresponds to a smaller ratio $\mu$ of equations/variables. More precisely, the probability that any vector is a solution of an equation is 1/2, hence it has probability $1/2^m$ of being a solution of the system. However, the number of regular vectors over $\F_2$ is only $l^w$. The mentioned threshold is thus obtained by equating
\[
\frac{l^w}{2^m}=1,
\]
which gives
\[
\mu =\frac{m}{n}=\frac{\log(l^w)}{n}=\frac{w\log(l)}{n}=\frac{\log(l)}{l}.
\]

Our analysis, especially the cryptanalysis through Gr\"obner basis computation, then mainly focuses on this case.

If instead we consider the same problem over a bigger field $\F_q$, then equating $((q-1)l)^w=q^m$ provides the determined case for \[
\mu=\frac{\log_q((q-1)l)}{l}.\]

\subsection{NP-completeness of the RMQ problem}
The decision problem of determining whether a boolean multivariate polynomial systems has solutions over $\F_2$ (respectively over its closure) is NP-complete (respectively NP-hard). This result has been proved in \cite{FY79} by reducing the 3-SAT problem to it. The result can be easily adapted to polynomials of degree 2, i.e. the decision MQ problem is NP-complete as well.
In the same work where it has been first introduced \cite{AFS05}, the regular syndrome decoding problem was also shown to be NP-complete. The polynomial-time reduction is an adaptation of the classic one from the Three Dimensional Matching problem \cite{BMT78}.

It is then not surprising that the decision RMQ problem (and consequently the search counterpart) is NP-complete, too. We provide here an elementary reduction from the MQ problem itself in the boolean case.
\begin{proposition}
    For any $l\ge 2$, the decision $l$-RMQ problem is NP-complete.
    \begin{proof}
        The decision $l$-RMQ problem is clearly in NP, as a candidate solution can be tested by 1) evaluating all the quadratic polynomials, and 2) checking the $l$-regularity. Both tasks can be done in polynomial time in the parameters.
        
        On the other hand, we prove the completeness by reducing the MQ problem to the $l$-RMQ problem in polynomial time. Let $S=\{f_1,\dots,f_m\}\subseteq \F_2[x_1,\dots,x_n]=R$ be a multivariate quadratic system. Let \\$R'= \F_2[x_{1,1},\dots,x_{1,l},\dots,x_{n,1},\dots,x_{n,l}]$ and consider the natural embedding $\phi\colon R \xhookrightarrow{} R'$ that sends $x_i$ to $x_{i,1}$. Let $f_j'\eqdef \phi(f_j)$ for all $j\in\Iintv{1}{m}$ and $S'=\{f_1',\dots,f_m'\}\subseteq R'$.

        Now let $\vv=(v_1,\dots,v_n)$ be a solution for the system associated with $S$. Define 
    \[
      \vv'\eqdef (v_1,v_1+1,0,\dots,0,v_n,v_n+1,0,\dots,0)\in \F_2^{nl}.
    \]
        Then, for any $j\in\Iintv{1}{m}$, $f_j'(\vv')=f(\vv)=0$, i.e. $\vv'$ is a solution for the system associated with $S'$. Moreover $\vv'$ is $l$-regular, since in each block
        \[
        (v_i,v_i+1,0,\dots,0)
        \]
        exactly one entry (one between the first and the second) is nonzero.
        Vice versa, let $\vv'= (v_{1,1},\dots,v_{1,l},\dots,v_{n,1},\dots,v_{n,l})$ be a regular solution for $S'$ and $\vv \eqdef (v_{1,1},\dots,v_{n,1})\in \F_2^n$. Then, by construction of $f_j'$, $f_j(\vv)=f_j'(\vv')=0,$ for any $j\in\Iintv{1}{m}$ which implies that $\vv$ is a solution for $S$. This shows that the decision MQ problem for $S$ and the decision RMQ problem for $S'$ have either both positive answer or both negative answer. Since $S'$ can be constructed in polynomial time from $S$, the reduction is polynomial-time.
    \end{proof}
    \begin{remark}
        The simple reduction provided introduces a linear blow-up in the window length $l$. This, however, is not indicative of how the hardness of the MQ problem diverges when adding the regular constraint and the blow-up can probably be improved. The next subsection shows that in the regime of unique solution, the exhaustive search changes only by a factor $l/\log(l)$ in the exponent of the complexity. The analysis of more advanced methods for the average-case problem will then appear in the following sections.
    \end{remark}
\end{proposition}

\subsection{Exhaustive search}
The most naive approach for solving an algebraic over a (small) finite field is to evaluate it over all elements in the vector space until a zero that is common to all equations is found. For standard multivariate quadratic systems the evaluation can be performed in an extremely efficient way using Gray codes \cite{BCCCNSY10} and the exponential cost is just given by the enumeration of all possible vectors: $\mathcal{O}^*(2^n/n_{\text{sol}})$, where $n_{\text{sol}}$ is the number of solutions.

We do not focus on the matter of adapting the efficient solution space enumeration to the regular setting, as the brute force approach only serves here as a benchmark for other algorithms. Admitting an efficient adaptation of the enumeration, the cost of such an approach in the regular framework amounts to $\mathcal{O}^*(l^w/n_{\text{sol}})$. In the case of a ``randomly'' sampled system with $m=\mu n$, $\mu=\frac{\log(l)}{l}$, equations, the complexity becomes
\[
\mathcal{O}^*(2^{\mu n})=\mathcal{O}^*\left(2^{\frac{\log(l)}{l} n}\right).
\]
More in general, the brute force approach over the field $\F_q$ has complexity 
\[
\mathcal{O}^*\left(2^{\frac{\log((q-1)l)}{l} n}\right).
\]

\section{Algebraic Cryptanalysis}
Besides brute force, the most common tool to solve a polynomial system and analyze the corresponding complexity is to run the XL algorithm/compute a Gr\"obner basis for the associated ideal. We first explain how to model the regular constraint with algebraic equations and then study relevant objects like the Hilbert series of the associated ideal and the degree of regularity to derive an estimation of the complexity of solving the system. Since we focus on the case of boolean systems, we also make use of and adapt in several ways the hybrid strategy to improve the algorithm.

\subsection{The algebraic modeling}
In addition to the initial system of equations obtained by equating to 0 the quadratic polynomials in $S_{\mathsf{init}}=\{P_i(\xv) \mid i\in \Iintv{1}{m}\}$, we first refine the ideal by adding add the $n$ polynomials corresponding to the field equations over $\F_2$
\[
S_{\mathsf{FE}}\eqdef \{ x_{i,j}^2-x_{i,j} \mid i\in \Iintv{0}{w-1}, j \in \Iintv{1}{l}\}.
\] 
which force solutions found through the Gr\"obner basis computation to lie over $\F_2$ instead of over its closure.

The weight constraint of each window is modeled by two families of equations:
\begin{itemize}
    \item The following $w\binom{l}{2}$ \textit{homogeneous} quadratic polynomials express the fact that each window has at most a nonzero entry
\[
    S_{\mathsf{Q}}\eqdef \{ x_{i,j_1} x_{i,j_2} \mid 0\le i\le w-1, 1\le j_1<j_2\le l\}.    
\] 
We have seen that the system $S_{\mathsf{init}}$ could be taken in a way that the quadratic monomials of the form $x_{i,j_1} x_{i,j_2}$ already miss from all the equations. However, the constraints from $S_{\mathsf{Q}}$ are still necessary to remove monomials of higher degree divisible by such $x_{i,j_1} x_{i,j_2}$'s during the Gr\"obner basis computation.
\item The following $w$ \textit{affine} linear polynomials express the fact that the sum (over $\F_2$) of the entry in each window is 1 (i.e. the weight of each window is odd)
\[
    S_{\mathsf{L}}\eqdef \left\{ \sum_{j=1}^l x_{i,j}+1 \mid i\in\Iintv{0}{w-1}\right\}.
\]
Note that each linear equation enables the elimination of one variable, so in practice we can set up a system with only $n-w=(l-1)w$ variables and recover the values of the $w$ eliminated variables by substitution once the system has been solved.
\end{itemize}
Since the only positive odd number smaller than 2 is 1, the two systems $S_{\mathsf{Q}}$ and $S_{\mathsf{L}}$ together model exactly the regular constraint of the solution. The equations from $S_{\mathsf{Q}}$ alone instead would model the ``at most regular'' constraint.
One can see that the equations in $S_{\mathsf{Q}}$ and $S_{\mathsf{L}}$ are exactly the same as those analyzed in \cite{BO23} for the regular syndrome decoding problem. 

The system that models the RMQ problem over $\F_2$ is then the following.
\begin{modeling}[Over $\F_2$] \label{modeling: F2}
    Given a system $S_{\mathsf{init}}=\{P_i(\xv) \mid i \in \Iintv{1}{m}\}$ of quadratic equations over $\F_2$, the algebraic system that models the RMQ problem over $\F_2$ is given by
    \begin{itemize}
    \item the system of $m+n+w\binom{l}{2}$ quadratic and $w$ linear equations corresponding to the polynomials
    \[
S\eqdef S_{\mathsf{init}} \cup S_{\mathsf{FE}} \cup S_{\mathsf{Q}} \cup S_{\mathsf{L}}
\]
\item in the $n=lw$ variables $x_{i,j}$'s.
    \end{itemize}
\end{modeling}
More in general, we define an algebraic modeling for big fields, where the field equations are not used and the nonzero entries are not necessarily equal to 1.
\begin{modeling}[Over a big field $\F_q$] \label{modeling: Fq}
    Given a system $S_{\mathsf{init}}=\{P_i(\xv)=0 \mid i \in \Iintv{1}{m}\}$ of quadratic equations over $\F_q$, the algebraic system that models the RMQ problem over $\F_q$ is given by
    \begin{itemize}
    \item the system of $m+w\binom{l}{2}$ quadratic equations corresponding to the polynomials
    \[
S\eqdef S_{\mathsf{init}} \cup S_{\mathsf{Q}}
\]
\item in the $n=lw$ variables $x_{i,j}$'s.
    \end{itemize}
\end{modeling}
These are the modelings for the plain Gr\"obner basis computation, we will see afterward several choices for the hybrid approach. While the former can be analyzed for big fields, the latter are relevant for small fields only and we will then restrict the hybrid methods analysis to boolean functions.

\subsection{The general proof framework for the Hilbert series derivation}
For each plain or hybrid strategy described in this section, we provide the Hilbert series of the associated (possibly specialized) ideal. Instead of providing the proof in each case separately, we give here a general framework of the arguments that lead to these results and devote the space of each subsection to the computation of the asymptotic degree of regularity.
The reasons behind this choice are twofold:
\begin{enumerate}
    \item At a macro level, all these proofs follow the same pattern with just a slight adaptation of the initial assumption.
    \item The full proof details have already been given in \cite{BO23}. In particular, the structured part of the system is exactly the same as for the regular syndrome decoding problem and we just recall the results holding for the structured system. The impact of the instance-specific quadratic equations can then be analyzed analogously to parity-check equations for the regular syndrome decoding problem according to a more general framework.
\end{enumerate}

We then proceed by briefly recalling each step of this analysis.
\subsubsection{Structured part}
The structured (and fixed) part of the system is $S_{\mathsf{fix}}=S_{\mathsf{Q}}$ for big fields and $S_{\mathsf{fix}}=S_{\mathsf{Q}}\cup S_{\mathsf{FE}}\cup S_{\mathsf{L}}$ in the boolean case.
The Hilbert series of the high-degree part of the structured subsystem alone can be derived as follows.
\begin{lemma}[from Lemma 6, \cite{BO23}]
    Let $R=\F_2[\xv]$. Let $S_{\mathsf{L}}^{(h)}=  \left\{ \sum_{j=1}^l x_{i,j} \mid i\in\Iintv{0}{w-1}\right\}$ and $S_{\mathsf{FE}}^{(h)}= \{ x_{i,j}^2 \mid i\in \Iintv{0}{w-1}, j \in \Iintv{1}{l}\}.$ be the part of highest degree of $S_{\mathsf{L}}$ and $S_{\mathsf{FE}}$ respectively. Then
\[
\HS_{R/\langle S_{\mathsf{Q}}\cup S_{\mathsf{L}}^{(h)} \cup S_{\mathsf{FE}}^{(h)}\rangle}(z)= \left(1+(l-1)z\right)^w.
\]
\end{lemma}
\begin{lemma}[Lemma 4, \cite{BO23}]
Let $R=\F[\xv]$, with $\F$ a finite field. Then
\[
\HS_{R/\langle S_{\mathsf{Q}}\rangle }(z)= \left(1+l\frac{z}{1-z}\right)^w=\left(\frac{1+(l-1)z}{1-z}\right)^w.
\]
\end{lemma}

Despite not rewriting the proofs, we just recall that the series is obtained by first determining the Hilbert series ``for one block'' as $1+(l-1)z$ or $1+l\frac{z}{1-z}$ and then applying a standard result about how the Hilbert series (more generally the generating function) compose with Cartesian product \cite{FS09}. Essentially, as the fixed part does not introduce any constraint between variables of different windows, the whole Hilbert series is obtained by multiplying the series of each block. This argument allows to determine the Hilbert series of a specialized system, where some of the variables have been fixed to 0. The most general specialization that we analyze is when any window potentially has a different number of entries set to 0. 
In particular, we fix to 0 $l-l'$ variables (wlog $x_{i,1}=\dots=x_{i, l-l'}=0$) for each window of a relative portion of $\gamma_{l'} \ge 0$ windows, for any $l'\in \Iintv{1}{l}$ in such a way that $\sum_{l'=1}^{l} \gamma_{l'} =1$ and we call $S_{\mathsf{guess}}$ the set of variables, which can be seen as homogeneous linear polynomials, that are fixed to 0. The same arguments used in \cite{BO23} and just recalled here readily imply, in the boolean case, 
\[
\HS_{R/\langle S_{\mathsf{Q}}\cup S_{\mathsf{L}}^{(h)} \cup S_{\mathsf{FE}}^{(h)}\cup S_{\mathsf{guess}}\rangle }= \prod_{l'=1}^{l} (1+(l'-1)z)^{\gamma_i w}.
\]

Adding the specialized equations to the structured part allows to unify all cases in a single assumption.

\subsubsection{Assumptions mimicking semi-regularity}
The computation of the Hilbert series for the entire system, i.e. including the non-structured part $S_{\mathsf{init}}$, is based on the  assumption that acts as that of semi-regularity, but with respect to the quotient ideal of the fixed part.
\begin{assumption} \label{assumption: semi-regular}
    Consider an instance $S_{\mathsf{init}}=\{1,\dots,f_m\}$ of Modeling~\ref{modeling: F2} or Modeling~\ref{modeling: Fq} and let $d_{reg}$ be the degree of regularity of $\mathcal{I}=\langle S_{\mathsf{init}}\rangle$. Define the quotient ring $\cQ \eqdef R/\langle S_{\mathsf{Q}}\cup S_{\mathsf{L}}^{(h)} \cup S_{\mathsf{FE}}^{(h)}\rangle$ in the non-hybrid boolean case, $\cQ \eqdef R/\langle S_{\mathsf{Q}}\cup S_{\mathsf{L}}^{(h)} \cup S_{\mathsf{FE}}^{(h)} \cup S_{\mathsf{guess}}^{(h)} \rangle$ in the hybrid boolean case or $\cQ \eqdef R/\langle S_{\mathsf{Q}} \rangle $ in the (non-hybrid) big field case. We assume that for $i\le i\le m$, $g_i f_i^{(h)}=0$ in $\cQ/\langle f_1^{(h)},\dots,f_{i-1}^{(h)}\rangle$ with $\deg(g_i f_i^{(h)})<d_{reg}$ implies $g_i=0$ in $\cQ/\langle f_1^{(h)},\dots,f_{i-1}^{(h)}\rangle$.
\end{assumption}

We let the reader appreciate the strong similarity of this assumption with with Definitions~\ref{def: semireg} and \ref{def: semiregF2}. We recall that analogous hypotheses has been used in \cite{BO23} for the regular syndrome decoding problem, by just replacing the quadratic equations defining the instance with the parity-check equations. Moreover, the assumption, formalizing the fact that randomly generated quadratic polynomials usually behave nicely in the quotient ring with respect to the structural part, has been tested extensively for several parameters.

\subsubsection{The random quadratic equations on top of the structured part}
Retrieving the Hilbert series of $R/\langle S^{(h)} \rangle$ from the quotient $\cQ$ of Assumption~\ref{assumption: semi-regular} is simply a particular case of \cite[Sections 3.3.1, 3.3.2]{B04}, cf. also \cite{BO23}. In particular, for each quadratic polynomial in $S_{\mathsf{init}}$, the Hilbert series is multiplied by $\frac{1}{1+z^2}$ in the boolean case and by $1-z^2$ in the big field case.
Moreover, the system is assumed to behave ``semi-regularly'' with respect to the extra variable $y$ used to homogenize it (cf. \cite[Section 4.3]{BO23}), which provides a factor $\frac{1}{1-z}$. Finally, the series obtained has to be truncated at the first non-positive coefficient.

All the Hilbert series proposed in the next subsections can thus be determined in a straightforward way within this framework.

\subsection{Plain Gr\"obner basis analysis}

Let $\xv'\eqdef (x_1,\dots,x_n,y),$ $R\eqdef\F_2[\xv']$ and $\mathcal{I}$ be the polynomial ideal generated by $S^{(y)}$.
The Hilbert series of $R/\mathcal{I}$ is
\[
\HS_{R/\mathcal{I}}(z)=\left[\frac{(1+(l-1)z)^w}{(1-z)(1+z^2)^m}\right]_{+},
\]
where the term $(1+z^2)^m$ accounts for the initial $m$ quadratic equations and the term $1-z$ for the homogenization variable $y$. 

We make use of complex analysis techniques, namely the saddle-point method, to estimate asymptotically the degree of regularity of the system from its Hilbert series. We refer the reader to \cite{W01} for a broader general discussion of these techniques. These methods have already been widely used in the context of studying the Hilbert series. Two examples that are close to our analysis are \cite{B04} for classic semi-regular systems and \cite{BO23} for the regular syndrome decoding modeling. Here we provide directly the formula for computing the asymptotic degree of regularity (and consequently the complexity) for the plain Gr\"obner basis approach over $\F_2$. All the necessary computation and explanations to get this value are moved to Appendix~\ref{app: plainF2}.

The asymptotic relative degree of regularity is the smallest positive root $\bar{\delta}$ of the quartic polynomial
\begin{equation*}
r_4 \delta^4+ r_3 \delta^3+r_2\delta^2+r_1\delta +r_0
\end{equation*}
in the variable $\delta$, whose coefficients are polynomial functions of $\mu$ and $l$:
\[
\begin{cases}
    r_4 &= ((l-1)^2+1)^2 \\
    r_3 &= 2\mu ((l-1)^2+1)((l-1)^2+3)-4\frac{(l-1)^2((l-1)^2+1)}{l}\\
    r_2 &= 4\mu^2 (2(l-1)^2+3)- 2\mu\frac{(l-1)^2(3(l-1)^2-1)}{l} +2\frac{(l-1)^2 (3 (l-1)^2 +1)}{l^2}\\
    r_1 &= 8 \mu^3+20\mu^2 \frac{(l-1)^2}{l} + 2\mu (l-1)^2 \frac{3(l-1)^2-5}{l^2 }-4 \frac{(l-1)^4}{l^3} \\
    r_0 &= -\mu^2 \frac{(l-1)^2}{l^2}-2\mu \frac{(l-1)^4}{l^3}+\frac{(l-1)^4}{l^4}.
\end{cases}
\]

Hence, the final complexity becomes
\[
\mathcal{O}^*\left(\binom{n}{\bar{\delta}n}^{\omega}\right)=\mathcal{O}^*\left(2^{\omega H(\bar{\delta})n}\right).
\]
A closed-form expression for roots of quartic polynomials exists but, as the coefficients are complicated functions in $l$ (and possibly $\mu$), we do not provide the exact formula. Here and after, we instead proceed by fixing $l$ and approximate the zeros of the polynomials to find $\bar{\delta}$. 
\begin{remark}
    To be more precise, the cost of reducing the Macaulay matrix would be $\binom{n}{\downarrow \bar{\delta}n}^{\omega}$. However, as it happens for the standard multivariate quadratic problem, $\bar{\delta}$ is small enough to have $\mathcal{O}^*\left(\binom{n}{\bar{\delta}n}^{\omega}\right)=\mathcal{O}^*\left(\binom{n}{\downarrow \bar{\delta}n}^{\omega}\right)$. The same approximation holds a fortiori for the next hybrid approaches, as $\bar{\delta}$ generally decreases with respect to plain Gr\"obner basis.
\end{remark}
\begin{remark}
    Due to the regular constraint and the filed equations, monomials whose degree with restricted to any block is at least 2 are reduced. Hence the total degree of a polynomial is at most $w$, which implies an upper bound on the relative degree regularity $\bar{\delta} \le w/n=1/l$ and consequently on the final complexity $\mathcal{O}^*\left(\binom{n}{\bar{\delta}n}^{\omega}\right)=\mathcal{O}^*\left(2^{\omega H(1/l)n}\right)$. It can be verified that for any $l$, the found degree of regularity correctly satisfies the upper bound, and it tends to it for $l\to \infty$.
\end{remark}
\subsubsection{Big fields}
For completeness, we conclude the section with an analysis of Modeling~\ref{modeling: Fq} for a big field $\F_q$. If $\mathcal{I}$ is the ideal generated by the polynomials in $S^{(y)}$, then the Hilbert series is
\[
\HS_{R/\mathcal{I}}(z)=\left[\frac{(1+(l-1)z)^w (1-z^2)^m}{(1-z)^{w+1}}\right]_{+}.
\]
As we did for the boolean case, we detail the computation needed to derive the asymptotic relative degree of regularity in Appendix~\ref{app: plainFq}. It turns out that this is the smallest positive root $\bar{\delta}$ of the quartic polynomial
\begin{equation*}
r_4 \delta^4+ r_3 \delta^3+r_2\delta^2+r_1\delta +r_0
\end{equation*}
in the variable $\delta$, whose coefficients are polynomial functions of $\mu$ and $l$:
\[
\begin{cases}
    r_4 &= 4l^2(2l+1)^2 \\
    r_3 &= -8\mu (l^3-4l^2-2l+4)-4(3l^2-8l+8)(l-2)\\
    r_2 &= -16\mu^2 (2l^2-4l-1)+ 8\mu(3l^3-14l^2+29l-24)+13l^2-48l+48\\
    r_1 &= -32\mu^3-16\mu^2 (5l-8) -4\mu (6l^2-23+24) -6(l-2) \\
    r_0 &= 4\mu^2 +4\mu(2l-3)+1.
\end{cases}
\]
Hence, the final complexity becomes
\[
\mathcal{O}^*\left(\binom{n+\bar{\delta}n}{\bar{\delta}n}^{\omega}\right)=\mathcal{O}^*\left(2^{\tau n}\right),
\]
with $\tau=\omega (1+\bar{\delta})H(\bar{\delta}/(1+\bar{\delta}))$.

For a fixed $q$, the plain Gr\"obner basis computation is convenient with respect to brute force for small enough values of $l$ and whose maximum (slowly) increases with $q$. To give some examples,
\begin{itemize}
    \item for $q\le 5$, brute force is always better;
    \item for $q\in \{7,8\}$ plain Gr\"obner basis beats brute force iff $l=2$;
    \item for $q\in \{31,32\}$ plain Gr\"obner basis beats brute force iff $l\le 8$;
    \item for $q\in \{255,256\}$ plain Gr\"obner basis beats brute force iff $l\le 31$.
\end{itemize}

\subsection{Hybrid full-windows guess}
A first idea of hybridization over $\F_2$ is to guess the nonzero entry of an amount of $\gamma w$ windows for some $\gamma \in [0,1]$. The choice $\gamma=0$ coincides with the plain Gr\"obner basis approach, while $\gamma=1$ with exhaustive search. The probability that the guess is consistent with the unique solution is $1/l^{\gamma w}$. The factor that takes into account the cost of guessing is $l^{\gamma w}.$

For a fixed specialization and after partial evaluation, this method reduces the initial RMQ problem to another RMQ problem with $(1-\gamma)w$ windows of length $l$ and hence with $(1-\gamma)n$ variables. The corresponding Hilbert series becomes
\[
\HS_{R/\mathcal{I}}(z)=\left[\frac{(1+(l-1)z)^{(1-\gamma)w}}{(1-z)(1+z^2)^m}\right]_{+}.
\]
By repeating the computation from the plain Gr\"obner basis strategy shown in Appendix~\ref{app: plainF2}, but replacing $w$ with $\gamma w$, we obtain the quartic  equation
$r_4 \delta^4+ r_3 \delta^3+r_2\delta^2+r_1\delta +r_0=0$
with
\[
\begin{cases}
    r_4 &= ((l-1)^2+1)^2 \\
    r_3 &= 2\mu ((l-1)^2+1)((l-1)^2+3)-4(1-\gamma)\frac{(l-1)^2((l-1)^2+1)}{l}\\
    r_2 &= 4\mu^2 (2(l-1)^2+3)- 2\mu(1-\gamma)\frac{(l-1)^2(3(l-1)^2-1)}{l} +2(1-\gamma)^2\frac{(l-1)^2 (3 (l-1)^2 +1)}{l^2}\\
    r_1 &= 8 \mu^3+20\mu^2(1-\gamma) \frac{(l-1)^2}{l} + 2\mu(1-\gamma) \frac{(l-1)^2 (3(l-1)^2-5)}{l^2 }-4 (1-\gamma)^3\frac{(l-1)^4}{l^3} \\
    r_0 &= -\mu^2 (1-\gamma)^2\frac{(l-1)^2}{l^2}-2\mu(1-\gamma)^3 \frac{(l-1)^4}{l^3}+(1-\gamma)^4\frac{(l-1)^4}{l^4}.
\end{cases}
\]
If $\bar{\delta}$ is the smallest positive root of this quartic equation, then the final complexity of the hybrid full-windows method is
\[
\mathcal{O}^*\left(l^{\gamma w}\binom{(1-\gamma)n}{\bar{\delta}n}^{\omega}\right)=\mathcal{O}^*\left(2^{\tau n}\right),
\]
with $\tau = \gamma\frac{\log(l)}{l}+\omega (1-\gamma)H(\frac{\bar{\delta}}{1-\gamma})$.
\subsection{Hybrid partial-windows guess}
Guessing the nonzero entry in a window can be equivalently seen as guessing the $l-1$ zero entries of the same window. Another possible way to hybridize over $\F_2$ is to guess only the same amount of zeros in each window, so that only $l'$ variables per block remain. This method contains the plain Gr\"obner basis one and the exhaustive search as subcases for $l'=l$ and $l'=1$ respectively. The probability that the guess is consistent with the unique solution is $(l/l')^w$. Hence, the factor that takes into account the cost of guessing is $(l'/l)^w$. Differently from the real number $\gamma$ from the previous hybridization, the value $l'$ only assumes discrete values.

For a fixed specialization and after partial evaluation, this method reduces the initial RMQ problem to another RMQ problem with $w$ windows of length $l'$ and hence with $l'w=\frac{l'}{l}n$ variables. The corresponding Hilbert series becomes
\[
\HS_{R/\mathcal{I}}(z)=\left[\frac{(1+(l'-1)z)^{w}}{(1-z)(1+z^2)^m}\right]_{+}.
\]
By repeating the computation from the plain Gr\"obner basis strategy shown in Appendix~\ref{app: plainF2}, but replacing $l$ with $l'$ (and paying attention to the fact that $w$ remains $1/l$ and not $1/l'$), we obtain the quartic  equation
$r_4 \delta^4+ r_3 \delta^3+r_2\delta^2+r_1\delta +r_0=0$
with
\[
\begin{cases}
    r_4 &= ((l'-1)^2+1)^2 \\
    r_3 &= 2\mu ((l'-1)^2+1)((l'-1)^2+3)-4(1-\gamma)\frac{(l'-1)^2((l'-1)^2+1)}{l}\\
    r_2 &= 4\mu^2 (2(l'-1)^2+3)- 2\mu(1-\gamma)\frac{(l'-1)^2(3(l'-1)^2-1)}{l} +2(1-\gamma)^2\frac{(l'-1)^2 (3 (l'-1)^2 +1)}{l^2}\\
    r_1 &= 8 \mu^3+20\mu^2(1-\gamma) \frac{(l'-1)^2}{l} + 2\mu(1-\gamma) \frac{(l'-1)^2 (3(l'-1)^2-5)}{l^2 }-4 (1-\gamma)^3\frac{(l'-1)^4}{l^3} \\
    r_0 &= -\mu^2 (1-\gamma)^2\frac{(l'-1)^2}{l^2}-2\mu(1-\gamma)^3 \frac{(l'-1)^4}{l^3}+(1-\gamma)^4\frac{(l'-1)^4}{l^4}.
\end{cases}
\]
If $\bar{\delta}$ is the smallest positive root of this quartic equation, then the final complexity of the hybrid partial-windows method is
\[
\mathcal{O}^*\left(\left(\frac{l}{l'}\right)^{w}\binom{(l'/l)n}{\bar{\delta}n}^{\omega}\right)=\mathcal{O}^*\left(2^{\tau n}\right),
\]
with $\tau=\frac{\log(l/l')}{l}+\omega \frac{l'}{l}H(\frac{\bar{\delta}l}{l'})$.

\begin{remark} \label{remark: l'=2}
    For all $4\le l<40 $, our estimates show that the choice $l'=2$ below provides the best trade-off. For $l\le \mathsf{split}=1000$, the optimal $l'$ never exceeds 3. More in general, the optimal $l'$ grows very slowly with respect to $l$, meaning the hybrid approach is quite unbalanced towards brute force rather than the plain Gr\"obner basis computation.
\end{remark}

\subsection{Hybrid different-windows guess}
The most general hybrid approach is to allow different partial guessing of zeros for each window, i.e. guessing $l-l'$ variables for each window of a relative portion of $\gamma_{l'} \ge 0$ windows, for any $l'\in \Iintv{1}{l}$ in such a way that $\sum_{l'=1}^{l} \gamma_{l'} =1$. 
The probability that the guess is consistent with the unique solution is $\prod_{l' \in \Iintv{1}{l}} (l/l')^{\gamma_{l'} w}=\prod_{l' \in \Iintv{1}{l-1}} (l/l')^{\gamma_{l'} w}$. Hence, the factor that takes into account the cost of guessing is $\prod_{l' \in \Iintv{1}{l-1}} (l'/l)^{\gamma_{l'} w}$. 
We remark that this hybridization can be seen as a natural generalization of all previous strategies:
\begin{itemize}
\item  $\gamma_{l}=1$ and $\gamma_{l'}=0$ otherwise, corresponds to a plain Gr\"obner basis computation;
\item $\gamma_{1}=1$ and $\gamma_{l'}=0$ otherwise, corresponds to brute force;
\item the minimum over the subset of tuples such that $\gamma_{1}=\gamma, \gamma_{l}=1-\gamma$ and $\gamma_{l'}=0$ otherwise, corresponds to hybrid full-window guess;
\item the minimum over the subset of tuples such that $\gamma_{\bar{l'}}=1$ for some $\bar{l'}$ and $\gamma_{l'}=0$ otherwise, corresponds to hybrid partial-window guess.
\end{itemize}
We also observe that, differently from the previous subcases, this additional layer of hybridization has not been considered in \cite{BO23} in the context of regular syndrome decoding.
Moreover, a generic specialization of this method does not reduce to another RMQ problem as windows can have different lengths.

To help the reader navigate through this general hybrid approach, we start by providing the case studies for different values of $l$, postponing the computation necessary to obtain the tuple $(\gamma_1,\dots,\gamma_l)$ that minimizes the complexity.

\subsubsection*{Full case study for $l=2$}
The regular multivariate quadratic problem with $l=2$ directly reduces to an instance of the standard multivariate quadratic problem. Indeed the linear equation $x-1+x_2=1$ from $S_{\mathsf{L}}$ (and corresponding to the first window) allows to eliminate the variable $x_2$ from the regular constraint $x_1 x_2=0$ from $S_{\mathsf{Q}}$, obtaining $x_1(x_1+ 1)=0$. But the latter equation is exactly the field equation for $x_1$, so the regular constraint is redundant. The same happens for any other window. Moreover, the complexity is the same as that of the ``hybrid partial-window guess'' approach, as one can only choose to guess either zero or one variable in a window. Since we are left with half the variables, the coefficient at the exponent gets exactly halved with respect to \cite{B04}, leading to a complexity 
\[
\mathcal{O}^*(2^{\frac{0.791}{2}n})=\mathcal{O}^*(2^{0.3955n}),
\]
corresponding to the vector $(\gamma_1,\gamma_2)=(0.449,0.551)$ and $\delta_{reg}=0.0154$, for $\mu=\frac{\log(l)}{l}=\frac{1}{2}$.

\subsubsection*{Full case study for $l=3$}
The case $l=3$ is therefore the first non-trivial example. The best choice of $\gamma$ with the different-windows hybrid approach is $(\gamma_1, \gamma_2, \gamma_3)\approx(0.127, 0.873, 0)$, leading to $\delta_{reg} \approx 0.01622$ and a complexity of 
\[\mathcal{O}^*(2^{0.4179n}).
\]
The fact that $\gamma_3=0$ means that at least 1 variable is thus guessed in each window and this results in finding the best trade-off for a random quadratic system in $m'=m$ unknowns and $n'=\frac{n}{3}$ variables. The classic hybrid approach applied for 
\[
\mu'=\frac{m'}{n'}=\frac{3m}{n}=3\frac{\log(l)}{l}=\log(3)
\]
is obtained for $\gamma\approx 0.127$, providing $\delta_{reg} \approx 0.0783$ and a complexity of $2^{0.6689n}$.

If we now multiply by the cost of guessing variables, we obtain
\[
\left(\frac{l}{l-1}\right)^{n/3}\cdot 2^{0.6689 n'}
\]
that leads to
\[\mathcal{O}^*\left(2^{\left(\frac{\log(3/2)}{3}+\frac{0.6689}{3}\right)n}\right)=\mathcal{O}^*(2^{0.4179 n})
\]
i.e. we retrieve exactly the complexity found with our method, thus corroborating and verifying our result.

\subsubsection*{Full case study for $l=4,5,6$}
For $l=4,5,6$, the best choice of $\gamma$ with the different-windows hybrid approach is $\gamma_2=1$ and $\gamma_{l'}=0$ otherwise, the RMQ problem is reduced to a random system of $m'=m$ equations in $n'=\frac{n}{l}$ variables. We thus have
\[
\mu'=\frac{m'}{n'}=\log(l).
\]
According to \cite{BFSS13}, whenever $0.55\mu'\ge 1$, the best algorithm is known to be a plain Gr\"obner basis computation (no hybrid). This is satisfied by any $l\ge 4$ and explains why, assuming $\gamma_{l'}$ for all $l'>2$, then $(\gamma_1,\gamma_2)=(0,1)$. We remark that, in these cases, the hybrid different-windows method boils down to the partial-windows one and it is very close to the brute force, as in each window only one variable less than the maximum is guessed. 

\subsubsection*{Large $l$}
For large (constant) values of $l$, estimating the best parameters for the hybrid different-windows approach becomes very costly. As already said in Remark~\ref{remark: l'=2}, if we restrict to the hybrid partial-window guess, the optimal choice seems to remain $\gamma_2=1$ and $\gamma_{l'}=0$ otherwise. We expect the result to change very little or nothing when allowing for different amounts of guesses in each window.

\subsubsection*{Hilbert series and general complexity}

The Hilbert series corresponding to the specialized systems is
\[
\HS_{R/\mathcal{I}}(z)=\left[ \frac{\prod_{l'=1}^{l} (1+(l'-1)z)^{\gamma_{l'} w}}{(1-z)(1+z^2)^m}\right]_{+}=\left[ \frac{\prod_{l'=2}^{l} (1+(l'-1)z)^{\gamma_{l'} w}}{(1-z)(1+z^2)^m}\right]_{+}.
\]
Consider the Cauchy integral
\[
\mathcal{I}_d(n)\eqdef \frac{1}{2\pi i}\oint \frac{1}{z^{d+1}} \frac{\prod_{l'=2}^{l} (1+(l'-1)z)^{\gamma_{l'} w}}{(1-z)(1+z^2)^m} dz 
\]
and set $f(z)$ such that $e^{nf(z)}= \frac{1}{z^{d+1}} \frac{\prod_{l'=2}^{l} (1+(l'-1)z)^{\gamma_{l'} w}}{(1+z^2)^m} $ (as the term $(1-z)$ is asymptotically negligible), i.e.
\[
f(z) \eqdef \left(\frac{1}{l}\sum_{l'=2}^{l}  \gamma_{l'} \ln(1+(l'-1)z)\right)-\mu\ln(1+z^2)-\delta\ln(z).
\]
We compute the saddle points from the function
\begin{align*}
zf'(z)=& \frac{1}{l}\left(\sum_{l'=2}^{l} \gamma_{l'} \frac{(l'-1)z}{1+(l'-1)z}\right)-\frac{2\mu z^2}{1+z^2}-\delta.
\end{align*}
For each tuple $(\gamma_{l'})_{l'}$ and a fixed window-length $l$, 
we multiply $zf'(z)$ by $1+z^2$ and all the denominators $1+(l'-1)z$'s such that $\gamma_{l'} \ne 0$, in order to obtain a polynomial 
\[g=\sum_{i=0}^{d_g} g_i z^i=\sum_{i=0}^{d_g} g_i(\delta) z^i
\]
in the variable $z$ and parameter $\delta$ of degree $d_g=1+\card{\{l' \mid 2\le l'\le l, \gamma_l' \ne 0\}}$, from which we can obtain the saddle points. These are given by the roots of the discriminant of $g$, whose degree is $d_g-2$. As long as at least two $\gamma_{l'}$ are nonzero, such a degree becomes $\ge 6$ and closed formula expressions for its root can not even be computed in theory. We recall that roots of the discriminant of $g$ are exactly the common roots of $g$ and its derivative $g'$, i.e. those of the resultant $\mathrm{Res}(g,g')$. Such resultant is a polynomial in $\delta$ of degree $d_g+(d_g-1)=2d_g-1$. Hence, the asymptotic value of the degree of regularity can be equivalently taken as the smallest positive root of the resultant:
\[
\bar{\delta}= \min \{\delta \in \R \mid (\mathrm{Res}(g,g'))(\delta)=0, \quad \delta >0\}.
\]

As the degree of the resultant potentially grows with $l$ and deriving a closed formula of its roots is not possible in general, we proceed by substituting all possible tuples $(\gamma_{l'})_{l'}$ according to some discrete split of the number of windows $\mathsf{split}$. Thus, a priori, we need to compute the complexity for 
\[
\binom{\mathsf{split}+l-1}{l-1}
\]
tuple values, which quickly increases with $l$. We remark that this discrete splitting is also useful for concrete parameters where the number does not tend to infinity and therefore taking $\mathsf{split}=w$ provides the best valid tuple $(\gamma_{l'})_{l'}$.

If $\bar{\delta}$ is the smallest positive root of the resultant, then the final complexity of the hybrid different-windows method is
\[
\mathcal{O}^*\left(\left(\prod_{l' \in \Iintv{1}{l-1}} \left(\frac{l}{l'}\right)^{\gamma_{l'} w}\right)\binom{\sum_{l' \in \Iintv{2}{l}}\gamma_{l'}\frac{l'-1}{l}n}{\bar{\delta}n}^{\omega}\right)=\mathcal{O}^*\left(2^{\tau n}\right),
\]
with $\tau=\sum_{l' \in \Iintv{1}{l-1}} \gamma_{l'}\frac{\log(l/l')}{l}+\omega \sum_{l' \in \Iintv{2}{l}}\gamma_{l'}\frac{l'-1}{l}H\left(\frac{\bar{\delta}\cdot l}{\sum_{l' \in \Iintv{2}{l}}\gamma_{l'}(l'-1)}\right)$.

\section{Cryptanalysis: the probabilistic polynomial method}
We now adapt the polynomial method introduced in \cite{LPTWY17} and then improved by \cite{BKW19, D20, D21} to the regular setting. 

We briefly and informally summarize the contributions brought by each article of this research line. Then, we provide a more detailed overview of their costs in standard case which serves as a comparison term for the subsequent improved adaptations to the regular setting. This family of algorithms, which are substantially different from algebraic approaches, relies on the strategy of representing a polynomial system through a single \textit{identifying polynomial}. The mathematical aspect of this idea is credited to Razborov and Smolensky \cite{R87, S87}. As we will recall, this approach is however impractical due to the very large degree attained by such polynomial. The probabilistic polynomial method consists in replacing the identifying polynomial with another one of much lower degree that represents the former with high probability.

This technique was previously developed for circuit design and its complexity analysis \cite{W14}, but used for the first time for polynomial system solving in \cite{LPTWY17}. The algorithm of Lokshtanov et al. is the first one that achieves an exponential speedup over exhaustive search in the worst case setting (for instance without semiregular or other algebraic assumptions and without requiring the system to be over- or underdetermined). The algorithm reduces the search version of the problem of solving a system to a linear number of calls to its decision counterpart and makes use of the probabilistic construction to solve the latter faster.

The first improvement on this first milestone is due to Bj\"orklund et al., leading to a further exponential speedup. The core idea of \cite{BKW19} is to isolate solutions in such a way that the decision version of the system solving problem can be in turn reduced to finding the parity of the number of solutions. We remark that in many cryptographic applications, algebraic systems are chosen in a way that only one solution exists. Our description thus immediately starts from Bj\"orklund et al.'s point of view. The other main contribution of \cite{BKW19} is to observe that a certain evaluation steps can be accelerated by using a self-recursion to smaller parity-counting instances. 

Building upon this version of the algorithm, Dinur remarked that the partial parities that need to be computed are strongly correlated with each other. Therefore, in \cite{D20} the multiple parity-counting problem is introduced, together with a procedure to solve it. By also making use of self-recursion, the algorithm form \cite{D20} attains the best asymptotic complexity for solving binary polynomial systems among all known methods.
Dinur later observed that the aforementioned reductions between problem add a large polynomial overhead that can be avoided completely. The same happens for the self-recursion, and \cite{D21} hence proposes a variant of the algorithm that has a worse exponential complexity but is better for concrete parameters of cryptographic interest because it cut out a significant polynomial factor. Differently from the previous approaches, \cite{D21} assumes that, for a given partition, an isolated solution exists with high probability. For cryptographic instances, this constraint is typically satisfied. 
We refer the reader to \cite{BBSV22} for a survey about this line of research and for an implementation and practical comparison of the mentioned methods.
In any case, as for the Gr\"obner basis approaches, our focus is mainly on the asymptotically relevant part and we provide the final complexity neglecting polynomial factors in $n$ (i.e. using the $\mathcal{O}^*$ notation).

\subsection{Preliminaries}
Given the initial system $S=\{P_1,\dots,P_m\}$, we define the polynomial
\[
F(\xv)=\prod_{i=1}^m (1+P_i(\xv)).
\]
We have that $\hat{\xv}$ is a solution of the system, if and only if $F(\hat{\xv})=1$. In order to simplify the discussion, we also assume that there is only one solution to the system. Indeed, the same assumption (which is common in cryptography) is made when adapting the method to the \textit{regular} framework, as the value $m$ is chosen in a way that the expected number of solutions is 1. The polynomial $F$ is thus called \textit{identifying polynomial} of $S$. However, the polynomial $F$ has a too high degree and the cost of evaluations becomes forbidding. Therefore, $F$ is replaced by a probabilistic polynomial $\tilde{F}$ constructed in the following way. Let $k<m$ be a positive integer. Pick uniformly at random a full rank matrix $\Am=(a_{i,j}) \in \F_2^{k\times m}$ and define for any $i\in\Iintv{1}{k}$ the quadratic polynomial
\[
R_i(\xv)=\sum_{j=1}^m a_{i,j} P_j(\xv).
\]
Finally, we define the probabilistic polynomial
\[
\tilde{F}(\xv)=\prod_{i=1}^k (1+R_i(\xv))
\]
which is the identifying polynomial of $\tilde{S}=\{R_1,\dots,R_k\}$ and whose degree is upper bounded by $2k$.
\begin{proposition}\label{prop: Pr_subsystem}
    For any $\hat{\xv} \in \F_2^n$, if $F(\hat{\xv})=1$, then $\tilde{F}(\hat{\xv})=1$. If instead $F(\hat{\xv})=0$, then $\Pr[\tilde{F}(\hat{\xv})=0]\ge 1-2^{-k}$.
\end{proposition}

The approach of \cite{BKW19} proceeds by reducing the problem of finding a solution to $S$ to a parity-counting problem. The first step consists of reducing the search version to the decision version of the problem, by fixing one by one the variables and calling $\Theta(n)$ times the decision algorithm.

Then the decision version of the problem is reduced to the parity-counting problem. As the solution is unique, the output of the decision algorithm is positive if and only if the output of the parity-counting algorithm is 1. In the case where the solution is not unique, \cite{BKW19} makes use of the Valiant-Vazirani affine hashing \cite{VV85}, which consists of adding random linear equations to the system in order to isolate a solution. In \cite{BKW19}, the parity of the number of solutions to $S$ is computed in parts by partitioning the length-$n$ vector of variables $\xv$ into 2 vectors $\yv=(y_1,\dots,y_{n-n_1})$ and $\zv=(z_1,\dots,z_{n_1})$, for some parameter $\gamma$ such that $n_1=\gamma n$, i.e.
\[
\sum_{\hat{x}\in \F_2^n} F(\hat{\xv})=\sum_{\hat{\yv}\in\F_2^{n-n_1}} \sum_{\hat{\zv}\in\F_2^{n_1}} F(\hat{\yv},\hat{\zv}). 
\]
When replacing $F$ with $\tilde{F}$, Proposition~\ref{prop: Pr_subsystem} and the union bound over all $\hat{\zv}\in \F_2^{n_1}$ provide the following upper bound on the partial parity-sums for each $\hat{\yv}\in \F_2^{n-n_1}$:
\begin{equation} \label{eq: rank_subsystem}
\Pr\left[\sum_{\hat{\zv}\in\F_2^{n_1}} \tilde{F}(\hat{\yv},\hat{\zv})=\sum_{\hat{\zv}\in\F_2^{n_1}} F(\hat{\yv},\hat{\zv})\right]\ge 1-2^{n_1-k}.
\end{equation}
The integer $k$ is thus chosen in an optimal way to obtain a probability non-negligibly larger than 1/2, that is $k=n_1+2$, i.e. $\lim_{n\to \infty} \frac{k}{n}=\gamma$.

A boolean function $F(\yv,\zv)$ can be written as
\[
F(\yv,\zv)=z_1\cdot\dots\cdot z_{n_1} F_1(\yv)+F_2(\yv,\zv),
\]
where at least a variable $z_i$ is missing in each monomial of $F_2(\yv,\zv)$. Then, because the field of coefficients is $\F_2$, the polynomial
\[
G(\yv)\eqdef \sum_{\hat{\zv}\in\F_2^{n_1}} \tilde{F}(\yv,\hat{\zv}) \in \F_2[\yv]
\]
is such that $G(\yv)=F_1(\yv)$. Therefore, the degree of $G(\yv)$ is at most
\[
d_G\le 2k-n_1.
\]
\subsubsection{Non-recursive version}
In order to interpolate $G(\yv)$ via the M\"obius transform, it is then enough to evaluate it on $\binom{[n-n_1]}{\downarrow d_G}\times \F_2^{n_1}$, instead of on $\F_2^n$. Each evaluation corresponds to finding solutions to $\tilde{S}$ via brute force, thus leading to a complexity of
\[
\mathcal{O}^*\left(\binom{[n-n_1]}{\downarrow d_G}2^{n_1}\right).
\]
The following step consists of evaluating $G(\yv)$ on all $\yv\in\F_2^{n-n_1}$ via M\"obius transform to obtain the partial parity sums with probability of correctness at least 3/4. Repeating $t=\Theta(n)$ times the procedure and applying a majority vote to choose the value of the partial parity leads to an error with exponentially small probability. This step trivially requires
\[
\mathcal{O}^*(2^{n-n_1})
\]
operations. Finally the algorithms outputs the parity as  $\sum_{\hat{y}\in \F_2^{n-n_1}} G(\hat{y})$.

Ignoring polynomial factors in $n$, the total complexity is
\[
\mathcal{O}^*\left(\binom{n-n_1}{\downarrow d_G}2^{n_1}+2^{n-n_1}\right).
\]
Let us define the function $g\colon [0,1]\to [0,1]$
\begin{equation}
    g(p) \eqdef (1-p)H^*\left(\frac{p}{1-p}\right).
\end{equation}
For a fixed $\gamma$, the asymptotic complexity can thus be written as
\[
\mathcal{O}^*(2^{\tau n}),
\]
where
\[
\tau \eqdef \max(g(\gamma)+\gamma, 1-\gamma).
\]
We want to minimize $\tau$ over $\gamma\in \Iintv{0}{1}$. Moreover, we observe that, whenever $\gamma\ge 1/3$, we have $g(\gamma)=1-\gamma$, therefore $\tau \ge 1$ does not improve upon brute force.
It turns out that the best choice to balance the two costs is $\gamma\approx 0.185 = 1-0.815$, which leads to a complexity of $\mathcal{O}^*(2^{0.815 n})$.

\subsubsection{Recursion \`a la Bj\"orklund et al.}
In \cite{BKW19}, a recursive self-reduction of the parity-counting problem is also used to improve the total complexity. Instead of evaluating over $\binom{[n-n_1]}{\downarrow d_G}\times \F_2^{n_1}$ for each element  $\yv \in \binom{[n-n_1]}{\downarrow d_G}$, the recursion is applied on the smaller space $\F_2^{n_1}$. The recursion stops after a constant number of iterations $D$. In the last step, an exhaustive search is run. A parity-counting instance thus reduces to many smaller instances of the same problem, where at each step the number of variables decreases by a factor $\frac{n_1}{n}=\gamma$. If we fix $\gamma$ and denote the complexity coefficient at depth $i$ of the recursion as $\tau_i$, the recursive formula thus gives
\begin{equation} \label{eq: bjorklund_recursive}
\tau_i= \gamma^i \max(g(\gamma)+\tau_{i+1}, 1-\gamma)\quad \forall i \in \Iintv{0}{D-1}
\end{equation}
and
\[
\tau_D=\gamma^D.
\]
We want to determine the value of $\gamma$ that optimizes $\tau_0$. By taking enough large $D$, \cite{BKW19} shows that
\[
\tau_0\le 1-\gamma \qquad \text{whenever} \qquad H^*\left(\frac{\gamma}{1-\gamma}\right)<1-\gamma,
\]
i.e.
\[
\tau_0 \le 1-\gamma \qquad \text{whenever} \qquad g(\gamma)<(1-\gamma)^2.
\]
In order to optimize the final complexity, the best choice consists of finding the value for which the increasing function $H^*\left(\frac{\gamma}{1-\gamma}\right)$ equals the decreasing function $1-\gamma$. In particular, we can select $\gamma\approx 0.197=1-0.803$, leading to a complexity of $\mathcal{O}^*(2^{0.803 n})$.

\subsubsection{Recursion \`a la Dinur}
In \cite{BKW19}, the many parity-counting instances are solved independently. However, this approach is suboptimal as such instances are related. Dinur introduces in \cite{D20} the \textit{multiple} parity-counting problem which solves all parity-counting instances at once. When recursion is applied, the multiple parity-counting instance is reduced to only a few smaller instances of the same problem. Moreover, the multiple counting algorithm is used to enumerate all the solutions of a polynomial system. 

 The multiple parity-counting instance must output the vector
 \[(G(\yv))_{\yv \in \binom{[n-n_1]}{\downarrow d_G}}.
 \]
 In addition to the parameter $\gamma= n/n_1$, a parameter $\eta$ is introduced in order to further split the space of the last $n_1$ variables into a right part of $n_2=(\gamma-\eta)n$ variables and a left part of $n_1-n_2=\eta n$ variables. In order to compute the multiple parity-counting instance, Dinur's algorithm applies a self-reduction to only one multiple parity-counting instance that outputs the vector whose coordinates are 
 \[
 G(\yv,\uv)=\sum_{\hat{\vv}\in \F_2^{n_2}} F(\yv, \uv, \hat{\vv}),
 \]
with $(\yv,\uv)\in \F_2^{n-n_1}\times \F_2^{n_1-n_2}$ such that the total weight is upper bounded by a certain quantity that depends on the rank $k'$ of the randomized subsystem that needs to be taken. Using the same argument derived from \eqref{eq: rank_subsystem} as before, we can choose $k'=n_2+2$, i.e. $\lim_{n\to \infty} \frac{k'}{n}= \gamma-\eta$, so that $\deg G(\yv,\uv)\le 2k'-n_2=n_2+4\sim n_2$. The following step is the interpolation of $G(\yv,\uv)$ with the M\"obius transform from its low-weight evaluations. The corresponding cost is 
\begin{equation} \label{eq: Dinur_intepolation}
\mathcal{O}^*\left(\binom{n-n_2}{\downarrow \deg G(\yv,\uv)}\right)=\mathcal{O}^*\left(\binom{n-n_2}{\downarrow n_2}\right).
\end{equation}

Finally $G(\yv,\uv)$ is evaluated on all the elements of 
\begin{equation} \label{eq: Dinur_eval}
\binom{[n-n_1]}{\downarrow \deg G(\yv)}\times \F_2^{n_1-n_2},
\end{equation}
so that the initial multi-parity counting output  
\[
(G(\yv))_{\yv \in \binom{[n-n_1]}{\downarrow d_G}}
 \]
can be reconstructed. The algorithm calls itself until depth $D$. If we denote with $n_1^{(i)}$ and $n_2^{(2)}$ the values of $n_1$ and $n_2$ at step $i\ge 0$ respectively, we obtain $n_1^{(0)}=n_1=\gamma n$ and $n_2^{(i)}=n_1^{(i+1)}\sim n_1^{(i)}-\eta n\sim n_1^{(0)}-i\eta n\sim (\gamma-i\eta)n$, for any $i\in \Iintv{0}{D-1}$, where $D$ is the integer such that $n_1^{(D)}\le \eta n$. Then we have $n_2^{(D)}=0$ and the algorithm runs brute force instead. 

We want to analyze the cost of each step of the recursion in order to find the time complexity of the whole algorithm. At each step $1\le i<D$, the cost \eqref{eq: Dinur_intepolation} coming from the interpolation can be written as
\[
\mathcal{O}^*\left(\binom{n-n_2^{(i)}}{\downarrow n_2^{(i)}}\right)=\mathcal{O}^*(2^{g(\gamma-i\eta)n}),
\]
while the evaluation step over \eqref{eq: Dinur_eval} takes time
\[
\mathcal{O}^*\left(\binom{n-n_1^{(i)}}{\downarrow n_1^{(i)}}2^{n_1^{(i)}-n_2^{(i)}}\right)=\mathcal{O}^*(2^{(\eta+g(\gamma-(i-1)\eta))n}).
\]
We then have to add the cost of brute force at depth $D$, which is
\[
\mathcal{O}^*\left(\binom{n-n_1^{(D)}}{\downarrow n_1^{(D)}}2^{n_1^{(D)}}\right)\le \mathcal{O}^*(2^{(\eta+g(\gamma-(D-1)\eta))n}).
\]
and the cost of the evaluation of $G(\yv)$ for all elements $\yv\in \F_2^{n-n_1}$, i.e.
\[
\mathcal{O}^*(2^{(1-\gamma)n}).
\]
We recall now how to choose the parameters that provide the total complexity claimed in \cite{D20}. We fix $\eta>0$ arbitrarily small, define $\tau=\max_{p\in [0,1/3]} g(p)=\log \psi \approx 0.6943$, where $\psi\approx 1.618$ is the golden ratio, and choose $\gamma=1-\tau$. In this way, each recursive step has an upper bound on the time complexity given by
\[
\mathcal{O}^*(2^{(\tau+\epsilon)n}).
\]
Moreover, since $n_2^{(D)}=0$, we derive that $D\eta\le\gamma\le (D+1)\eta$ and hence $D\sim \frac{\gamma}{\eta}=O(1)$ can be taken constant so that the final complexity remains
\[\mathcal{O}^*(2^{0.6943 n}).
\]

\subsection{Adapting to the regular framework}
The search to decision problem reduction can be easily transferred to the regular setting. Assuming a solution exists, for the first window all the $l$ possibilities are tested. Assuming the uniqueness of the solution, only one test gives a positive answer. Then the algorithm proceeds recursively with the next window. Thus, the number of calls needed is upper bounded by $wl=n$.


We define the polynomials $R_i$'s and $F$ in the same way. Proposition~\ref{prop: Pr_subsystem} still holds. However, since we want to find the unique \textit{regular} solution, we need to consider the parity of solutions over regular elements, i.e. $\sum_{\hat{\xv}\in \cR(\F_2^n)} F(\hat{\xv})=\sum_{\hat{\yv}\in \cR(\F_2^{n-n_1})} \sum_{\hat{\zv}\in\cR(\F_2^{n_1})} F(\hat{\yv},\hat{\zv})$ and we define
\[
G(\yv)=\sum_{\hat{\zv}\in\cR(\F_2^{n_1})} F(\yv,\hat{\zv}). 
\]

We want $\hat{\yv}\in \F_2^{n-n_1}$ to be an isolated solution, and we have
\[
\Pr\left[\sum_{\hat{z}\in \cR(\F_2^{n_1})} \tilde{F}(\hat{y},\hat{z})=\sum_{\hat{z}\in\cR(\F_2^{n_1})} F(\hat{y},\hat{z})\right]\ge 1-\frac{l^{\gamma w}}{2^k}.
\]
Therefore, we can pick $k=\floor{\gamma w \log l}+2$ so that the probability above remains above 1/2 by a constant factor. Asymptotically, this choice corresponds to \[k \sim \gamma w \log l.\] Let $\delta_{\vv}$ be the set of indices where $\vv$ is not zero, i.e.
\[
\delta_{\vv} \eqdef \{i \in \Iintv{1}{n} \mid v_i = 1\}.
\]
We can write the polynomial $F$ as
\[
F(\yv,\zv)= \sum_{\vv \in \cR(\F_2^{n_1})} \left(\prod_{i \in \delta_{\vv}} z_i\right) F_{\vv}(\yv)+F'(\yv,\zv),
\]
where each monomial appearing in $F'(\yv,\zv)$ misses completely at least one window of variables in $\zv$. Hence, if $l$ is even, then
\[
G(\yv)= \sum_{\zv\in\cR(\F_2^{n_1})} \left[ \sum_{\vv \in \cR(\F_2^{n_1})} \left(\prod_{i \in \delta_{\vv}} z_i\right) F_{\vv}(\yv)+F'(\yv,\zv) \right] = \sum_{\vv \in \cR(\F_2^{n_1})} F_{\vv}(\yv),
\]
and its degree is $d_G\le 2k - \gamma w\sim \gamma w(2\log(l)-1)$. If instead $l$ is odd, then $F'$ does not vanish when summing over because its monomials may appear an odd number of times. To circumvent this issue, we can guess a 0 position in each block and reduce the regular multivariate quadratic problem to a new instance of the new problem with $w$ windows of length $l-1$ (which is even). More in general, we can choose to initially fix an even number of zeros in each window when $l$ is even and an odd number of zeros in each window when $l$ is odd. In both cases, we are left with an even number $l'<l$ of variables for each window. From now on, we assume that $l$ is even, and take into account the cost of guessing the zeros in the final complexity. 

In the standard setting, the efficiency of the interpolation and evaluation steps is guaranteed by the fact the M\"obius transform allows to reconstruct the algebraic normal form (ANF) of a polynomial of degree $d$ from evaluations of vectors of weight at most $d$ only. In the regular setting, we can show that evaluations of regular vectors can be reconstructed from the knowledge of low-weight at most regular vectors. More precisely, we have the following results that is exploited throughout the algorithms to take advantage of the regular structure and improve on the complexity.
\begin{lemma}
    Let $\yv_1 \in \F_2^n$ and $\yv_2 \in \cAMR(\F_2^n)$, such that $\supp(\yv_1)\subseteq \supp(\yv_2)$. Then  $\yv_1 \in  \cAMR(\F_2^n)$.
    \begin{proof}
        In each window, $\yv_2$ has at most one coordinate equal to 1. Since $\supp(\yv_1)\subseteq \supp(\yv_2)$, the same holds for $\yv_1$. Therefore $\yv_1$ is also at most regular.
    \end{proof}
\end{lemma}
\begin{proposition}
    Let $F \colon \F_2^n \to \F_2$ be a boolean function of degree $d$. Let $\xv\in \cR(\F_2^n)$. Then the evaluations $F(\xv)$ can be computed from the evaluations of $F$ over all $\yv\in \cAMR\left(\binom{[n]}{\downarrow d}\right)$.
    \begin{proof}
        Let $S\subseteq \Iintv{1}{n}$ and $\xv_S$ be the monomial that is the product of all variables whose index belongs to $S$. We can thus write
        \[F(\xv)\eqdef \sum_{\card{S}\le d} a_S \xv^S.\]
        We first prove that from the evaluations of all at most regular vectors of weight at most $d$, we obtain the value of each $a_S$ such that $\xv_S$ is at most regular. We prove this first result by induction on the cardinality of $S$.
        
        \textbf{Base case:} if $\card{S}=0$, then $S=\emptyset$ and $a_0=F(\zerov)$ is known.
        
        \textbf{Inductive step:} Assume we know $a_{\hat{S}}$ for all $\hat{S}$ such that $\xv_{\hat{S}}\in \cAMR\left(\binom{[n]}{\downarrow w}\right)$. Let $S$ be such that $\xv_{S}\in \cAMR\left(\binom{[n]}{w+1}\right)$. We have
        \[
        F(\xv_S)=\sum_{\hat{S}\subseteq S} a_{\hat{S}}=a_S+\sum_{\hat{S}\subset S} a_{\hat{S}}.
        \]
        For any $\hat{S}\subset S$, by the previous lemma $\xv_{\hat{S}}\in \cAMR(\F_2^n)$, moreover $\hat{S}$ has weight at most $w+1-1=w$, hence  $\xv_{\hat{S}}\in \cAMR\left(\binom{[n]}{\downarrow w}\right)$ and  $a_{\hat{S}}$ is known. Therefore $a_S$ can also be computed.

        Now let $\xv_S\in \cR(\F_2^n)$. We have
        \[
        F(\xv_S)=\sum_{\hat{S}\subseteq S} a_{\hat{S}}=\sum_{\hat{S}\subset S, \card{\hat{S}}\le d} a_{\hat{S}},
        \]
        where the last equality follows from the fact that $\deg F =d$ and thus $a_{\hat{S}}=0$ whenever $\card{\hat{S}}>d$. Again by the previous lemma, all subsets $\hat{S}\subseteq S$ are such that $\xv_{\hat{S}}$ is at most regular. Hence $F(\xv_S)$ can be reconstructed because all the relevant $a_{\hat{S}}$'s are known.
    \end{proof}
\end{proposition}
If $2k-\gamma w \ge (1-\gamma) w$, we then need to compute $G(\yv)$ for all $\yv\in\cAMR( \F_2^{n-n_1})$, so that $G(\yv)$ can be interpolated. In this case, the parity counting computation already requires a number of operations given by $(l+1)^{(1-\gamma)w}l^{\gamma w} > l^w$ and is therefore worse than a brute force approach. We can thus restrict the (asymptotic) analysis to the case where
\begin{align*}
    & \gamma w (2\log(l) -1) < (1-\gamma) w\\
    \iff & \gamma < \frac{1}{2 \log(l)}.
\end{align*}
In such a case, the regular vectors $\yv$'s for which $G$ needs to be calculated are only those with total weight at most $d_G$, whose number is
\[
\sum_{i=0}^{2k-\gamma w} \binom{(1-\gamma)w}{i} l^i,
\]
Then $G(\yv)$ is interpolated from these values and finally evaluated on all the $l^{(1-\gamma)w}$ regular vectors. The final complexity for the non-recursive version of the algorithm thus becomes
\begin{equation} \label{eq: polynomial_regular}
\mathcal{O}^*\left(\left(\sum_{i=0}^{2k-\gamma w} \binom{(1-\gamma)w}{i} l^i\right)l^{\gamma w}+l^{(1-\gamma)w}\right).
\end{equation}

We now have to determine the relevant function $g: [0,1]\times Z_+ \to [0,1]$ (that depends on $\gamma$ and $l$) for the regular setting. It is defined as the function such that
\[
\mathcal{O}^*(2^{g(p, l)n})= \mathcal{O}^*\left(\sum_{i=0}^{p w (2\log(l)-1)} \binom{(1-p)w}{i} l^i\right).
\]
Since the sum on the right-hand side has a linear number of terms in $n$, the asymptotic is given by 
\[
\max_{i\in\Iintv{0}{\floor{p w (2\log(l)-1)}}} \binom{(1-p)w}{i} l^i.
\]
The argmax can be determined as the $i$ for which two consecutive terms are as close as possible. By imposing
\[
 1\approx\frac{\binom{(1-p)w}{i+1}l^{i+1}}{\binom{(1-p)w}{i}l^{i}}=\frac{l((1-p)w-i)}{i+1},
\]
we obtain
\[
i\sim \frac{l(1-p)w}{l+1}=\frac{(1-p)}{l+1}n.
\]
Therefore, whenever $p w (2\log(l)-1)\le \frac{l}{l+1}(1-p)w$, the maximum term is the last one, i.e.
\[
 \binom{(1-p)w}{p w (2\log(l)-1)} l^{p w (2\log(l)-1)}.
\]
Otherwise, we obtain
\[
2^{g(p, l)n}\sim \binom{(1-p)w}{\frac{(1-p)}{l+1}n}l^{\frac{(1-p)}{l+1}n}=(l+1)^{(1-p)w}.
\]
This leads to the definition
\[
g(p, l)\eqdef \begin{cases}
\frac{1}{l}\left((1-p) H\left(\frac{p(2\log(l)-1)}{1-p}\right)+p (2\log(l)-1) log(l)\right), & \text{if }p (2\log(l)-1)\le \frac{l}{l+1}(1-p)\\
    (1-p)\frac{\log(l+1)}{l},  & \text{otherwise.}
\end{cases}
\]
For fixed $l$ and $\gamma$, the asymptotic exponent coefficient of our algorithm becomes
\[
\max\left(g(\gamma, l)+\gamma\frac{\log(l)}{l}, (1-\gamma)\frac{\log(l)}{l}\right).
\]
The probability of guessing correctly $l-l'$ zeros in each window amounts to $(l/l')^w$ and the partial evaluations on the guessed zeros for each system equation can be done in time $\mathcal{O}^*(1)$. Therefore, the initial problem reduces in polynomial time to a new instance of regular multivariate quadratic with $w$ windows of length $l'$, so that the total number of variables is $n'=l'w=\frac{l'}{l}n$. Finally, when taking into account the initial guess of zeros in each window, the asymptotic complexity turns out to be
\[
\mathcal{O}^*(2^{\tau n })
\]
where
\[\tau \eqdef \min \left\{\frac{\log(l/l')}{l}+\frac{l'}{l}\max\left(g(\gamma, l')+\gamma\frac{\log(l')}{l'}, (1-\gamma)\frac{\log(l')}{l'}\right) \mid l' \text{even}, 0\le l' \le l, 0\le \gamma \le \frac{1}{2\log(l')}\right\}.
\]
\subsubsection{Recursion \`a la Bj\"orklund et al.}
We adapt now the recursion argument from \cite{BKW19} in the context of regular solutions. For a fixed $\yv$ of low weight, instead of partially evaluating $F(\yv,\zv)$ over all $\zv\in \cR(\F_2^{n_1})$ to obtain $G(\yv)$, many parity counting instances with a number of variables that decreases by a factor $\gamma$ are solved recursively. The term $l^{\gamma w}$ in \eqref{eq: polynomial_regular} is thus replaced with the cost of the recursive call. Hence, fixed $\gamma$ and called $\tau_i$ the cost at depth $i$, the recursive formula \eqref{eq: bjorklund_recursive} translates in the regular setting to
\begin{equation} \label{eq: Bjorklund_regularrecursive}
\tau_i= \gamma^i \max\left(g(\gamma, l)+\tau_{i+1}, (1-\gamma)\frac{\log(l)}{l}\right)\quad \forall i \in \Iintv{0}{D-1}
\end{equation}
and
\[
\tau_D = \gamma^D \frac{\log(l)}{l}.
\]
In order to find the optimal value for $\gamma$, in a way analogous to \cite{BKW19}, we prove the following result.
\begin{proposition}
    Let $0\le \gamma\le \frac{1}{2\log(l)-1}$ be fixed and $g(\gamma, l)\le(1-\gamma)^2 \frac{\log(l)}{l}$. Then 
    \[
\tau_0 \le (1-\gamma)\frac{\log(l)}{l}.
\]
\begin{proof}

We proceed by induction. For sufficiently big $D$, we have
\begin{align*}
\tau_{D-1}=&\gamma^{D-1} \max\left(g(\gamma, l)+\gamma_D\frac{\log(l)}{l}, (1-\gamma)\frac{\log(l)}{l}\right)\\=&\gamma^{D-1} \max\left((1-\gamma)^2 \frac{\log(l)}{l}+\gamma_D\frac{\log(l)}{l}, (1-\gamma)\frac{\log(l)}{l}\right)\\=&\gamma^{D-1}\frac{log(l)}{l} \max(1-2\gamma+\gamma^2+\gamma^D, 1-\gamma)\\=&\gamma^{D-1}(1-\gamma)\frac{log(l)}{l}.
\end{align*}
Let us now assume that $\tau_{i+1}\le \gamma^{i+1}(1-\gamma)\frac{log(l)}{l}$. The inductive step is to prove that 
\[\tau_{i}\le \gamma^{i}(1-\gamma)\frac{log(l)}{l},
\]
while the case $i+1=D-1$ shown above represents the base case.
From \eqref{eq: Bjorklund_regularrecursive},
\begin{align*}
\tau_i=& \gamma^i\max\left( \gamma^{i+1}(1-\gamma)\frac{log(l)}{l}+\tau_{i+1}, (1-\gamma)\frac{\log(l)}{l}\right)\\
\le& \gamma^i\max\left((1-\gamma)^2 \frac{\log(l)}{l}+\gamma^{i+1}(1-\gamma)\frac{log(l)}{l}, (1-\gamma)\frac{\log(l)}{l}\right)\\
=&\gamma^i(1-\gamma)\frac{log(l)}{l}\max((1-\gamma)+\gamma^{i+1}, 1)\\
=&\gamma^i(1-\gamma)\frac{log(l)}{l}.
\end{align*}
Hence, it follows by induction that  $\tau_0 \le (1-\gamma)\frac{\log(l)}{l}.$
\end{proof}
\end{proposition}
Depending on the initial amounts $l-l'$ of fixed zeros per block, the final complexity can thus be upper bounded by 
\[
\mathcal{O}^*(2^{\tau n})
\]
where
\[\tau \eqdef\min \left\{\frac{\log(l/l')}{l}+(1-\gamma_{l'})\frac{\log(l')}{l} \mid l' \text{even}, 0\le l' \le l\right\}.
\]
and
\[
\gamma_{l'}\eqdef \max\left\{\gamma \mid g(\gamma, l')\le(1-\gamma)^2 \frac{\log(l')}{l'}\right\}.
\]
We remark that from our experiments, the optimal value of $l'$ is always $l$ or $l-1$ depending on the parity of $l$, i.e. at most one 0 per window is guessed.

\subsubsection{Recursion \`a la Dinur}
Analogously to \cite{D20}, we define a multiple parity-counting problem that outputs
\[(G(\yv))_{\yv \in \cAMR(\F_2^{(1-\gamma)n})}.
 \]
 Then, we further split the block $\zv$ of length $n_1=\gamma n$ into two blocks $\uv$ and $\vv$ of length $n_1-n_2=\eta n$ and $n_2=(\gamma-\eta)n$ respectively. In this way, a self-reduction to a new multiple parity-counting instance is applied and such instance outputs the vectors with coordinates 
 \[
 G(\yv,\uv)=\sum_{\hat{\vv}\in \cR(\F_2^{n_2})} F(\yv, \uv, \hat{\vv}),
 \]
with $(\yv,\uv)\in \cAMR(\F_2^{n-n_1}\times \F_2^{n_1-n_2})$ of weight upper bounded by $\deg G(\yv,\uv)\le 2k'-\frac{2\log(l)-1}{l}n_2\sim  (\gamma-\eta)\frac{2\log(l)-1}{l} n$. 
The interpolating step through the M\"obius transform has asymptotic cost
\begin{equation}
\mathcal{O}^*(2^{g(\frac{n_2}{n}, l)n}).
\end{equation}

Finally $G(\yv,\uv)$ is evaluated on all the elements of 
\begin{equation}
\cAMR\left(\binom{[n-n_1]}{\downarrow \deg G(\yv)}\right)\times \cR(\F_2^{n_1-n_2}),
\end{equation}
so that the initial multi-parity counting output  
\[
(G(\yv))_{\yv \in \binom{[n-n_1]}{\downarrow d_G}}
 \]
can be reconstructed. 

The same reasoning exhibited before allows to evaluate the cost of each recursive call. 

With the notation already introduced, at each step $1\le i<D$, the cost of interpolation can be written as
\[
\card{\cAMR\left(\binom{[n-n_2^{(i)}]}{\downarrow n_2^{(i)}/l}\right)}=\mathcal{O}^*(2^{g(\gamma-i\eta, l)n}),
\]
while the evaluation step takes time
\[
\card{\cAMR\left(\binom{[n-n_1^{(i)}]}{\downarrow n_1^{(i)}/l}\right)}\cdot \card{\cR(\F_2^{n_1^{(i)}-n_2^{(i)}})}=\mathcal{O}^*(2^{(\eta\frac{\log(l)}{l}+g(\gamma-(i-1)\eta, l))n}).
\]
We then have to add the cost of brute force at depth $D$, which is
\[
\card{\cAMR\left(\binom{[n-n_1^{(i)}]}{\downarrow n_1^{(i)}/l}\right)}\cdot \card{\cR(\F_2^{n_1^{(i)}-n_2^{(i)}})}\le \mathcal{O}^*(2^{(\eta\frac{\log(l)}{l}+g(\gamma-(D-1)\eta, l))n}).
\]
and the cost of the evaluation of $G(\yv)$ for all elements $\yv\in \cR(\F_2^{n-n_1})$, i.e.
\[
\mathcal{O}^*\left(2^{(1-\gamma)\frac{\log(l)}{l}n}\right).
\]
By taking into account the initial guess of zeros, the complexity exponent coefficient then becomes
\[
\mathcal{O}^*(2^{(\tau+\eta\frac{\log(l')}{l})n}).
\]
where
\[
\tau \eqdef \min\left\{\max\left\{\frac{\log(l/l')}{l}+\frac{l'}{l}g(\gamma, l') \mid \gamma \in [0,\hat{\gamma}_{l'}] \right\} \mid  l' \text{even}, 0\le l' \le l \right\}.
\]
and
\[
\hat{\gamma}_{l'} \eqdef \min \left\{\gamma \mid g(\gamma, l') \ge (1-\gamma)\frac{\log(l')}{l'}\right\}.
\]
Again, the term $\eta$ can be chosen arbitrarily small (and we thus neglect it in the comparison with other approaches) while keeping constant the number of iterations $D$. Moreover, as well as for the recursion \`a la Bj\"orklund et al., the optimal value of $l'$ is either $l$ or $l-1$, depending on the parity of $l$.

\section{An alternative interpretation of boolean regular multivariate quadratic systems}
Despite each entry of the solution vector can assume two values, it is evident that the information content/entropy carried out by the corresponding variable is lower than that for a random system with no constraints on the solution. It is then natural to wonder whether the solution can be written with less bits of information while preserving the possibility of expressing it as the solution of an algebraic system. We present here an alternative way to describe an instance of the regular multivariate quadratic problem that uses less variables, indeed. Let us consider a \textit{binary} multivariate quadratic system whose solution of interest is regular and with the usual parameters. For the sake of simplicity, we restrict to the case where the window length is a power of 2, i.e. $l=2^s$ for some positive integer $s$. The same description could be carried out to the more general case with a slightly suboptimal algebraic system.

We observe that the set of variables of $\F_2[x_1,\dots,x_l]$ can be mapped injectively into the set of monic polynomials of degree $s=\log(l)$ in $\F_2[x_1',\dots,x_s']$. A canonical example of such map is
\[g \colon x_j \mapsto \prod_{a=1}^s (x_a'+\mathsf{bin}_a(j-1)),\]
where $\mathsf{bin}_a(b)$ denotes the $a$-th least significant bit of the binary representation of $b$. Then, $\bar{g}$ induces a bijection between the set of binary vectors of length $l$ and weight 1 and the vector space $\F_2^{s}$: $\vv$ is sent to the unique vector $\vv' \in \F_2^{s}$ such that for all $i\in \Iintv{1}{l}$,
\[
\ev_{x_i}(\vv)=v_i=\ev_{g(x_i)}(\vv'),
\]
where $\ev_{P(\xv)}(\vv)$ is the evaluation $P(\vv)$.

The map $g$ extends via multiplication to monomials (including those formed by variables from different blocks). We split the $\xv'$ vector into $w$ blocks of length $s$ and denote its entries with two indexes, so that we define
\[
\bar{g} \colon  \prod_{i=0}^{w-1}\prod_{j=1}^{l} x_{i,j}^{\alpha_{i, j}} \mapsto  \prod_{i=0}^{w-1} \prod_{j=1}^{l} \prod_{a=1}^s (x_{i,a}'+\mathsf{bin}_a(j-1))^{\alpha_{i,j}}.
\]
We are interested in evaluation of regular binary vectors, for our purposes it is thus enough to restrict the map $\bar{g}$ to polynomials of degree at most 1 in each block of variables:
\begin{equation} \label{eq: bar_g}
\bar{g} \colon  \prod_{i=0}^{w-1} x_{i,j_i} \mapsto  \prod_{i=0}^{w-1} \prod_{a=1}^s (x_{i,a}'+\mathsf{bin}_a(j_i-1)).
\end{equation}
If we now consider an instance of the multivariate quadratic problem, the map $\bar{g}$ sends, by linearity, any multivariate quadratic equation into an equation of degree $2s$ in $sw$ variables. We thus reduced the multivariate quadratic problem to a multivariate degree-($2s$) problem in less variables. Notice that even if the instance of the RMQ problem is picked randomly, the new system obtained still preserves some structure. Indeed the only terms that can appear are those coming from the expansion of
\begin{equation} \label{eq: terms_prime}
\prod_{a=1}^s (x_{i_1,a}'+b_{i_1,a})(x_{i_2,a}'+b_{i_2,a}),
\end{equation}
for some $b_{i_1,a},b_{i_2,a}\in \F_2$ and $i_1\neq i_2$.
We will discuss how this structure affects the Hilbert series of the associated ideal and consequently the resolution of the system via Gr\"obner basis techniques later. On the other hand, the threshold ratio of equations over variables that corresponds to the unicity of the solution becomes
\[
\mu'=\frac{m'}{n'}=\frac{m}{sw}=1.
\]
This value resembles that of random systems and reflects the fact that no conditions are put on the shape of the solution anymore.
We remove the prime symbol from now on.

\subsection{Algebraic cryptanalysis of the alternative modeling}
We want to study the Hilbert series of the homogenization of the system obtained from the construction above. We call $\cI$ the ideal generated by it. First of all, we characterize the admissible terms in the Macaulay matrix at a given degree $d$. After homogenization, the only terms that can appear in a degree-($2s$) equation can be derived by homogenizing \eqref{eq: terms_prime}:
\begin{equation*} 
\prod_{a=1}^s (x_{i_1,a}+b_{i_1,a}y)(x_{i_2,a}+b_{i_2,a}y)
\end{equation*}
and then expanding it, i.e.
\[\left\{\left(\prod_{a_1 \in A_1} x_{i_1,a_1} \cdot \prod_{a_2 \in A_2 } x_{i_2,a_2}\right)
 y^{2s-\card{A_1}-\card{A_2}} \mid i_1\neq i_2, A_1,A_2\subseteq \Iintv{1}{l}\right\}.\]

The characterization of admissible terms in the Macaulay matrix at a given degree $d\ge 2s$ boils down to listing all the terms that are contained in the ideal generated by the monomials above. It is straightforward to see that a monomial 
\[
 \prod_{i=0}^{w-1}\prod_{j=1}^{l} x_{i,j}^{\alpha_{i, j}}\cdot y^{\alpha_h}
\]
can appear if and only if 
\[
\exists i_1\neq i_2 \text{ s.t. }\left(\sum_{j=1}^l  \alpha_{i_1,j}+\alpha_{i_2,j}\right)+\alpha_h \ge 2s.
\]
Let $\mathcal{M}_{NA}(d)$ be the set of non-admissible monomials at degree $d$,
\[
\mathcal{M}_{NA}(d)\eqdef \left\{ \prod_{i=0}^{w-1}\prod_{j=1}^{l} x_{i,j}^{\alpha_{i, j}}\cdot y^{\alpha_h} \mid \sum_{i,j} \alpha_{i,j}+\alpha_h =d \land  \forall i_1\neq i_2, \left(\sum_{j=1}^l  \alpha_{i_1,j}+\alpha_{i_2,j}\right)+\alpha_h < 2s\right\}.
\]
We are mainly interested in determining the maximum degree at which such set are not empty, because this implies a corresponding non-zero Hilbert function.
\begin{proposition}
    For any $d> (s-1)w+1$, $\mathcal{M}_{NA}(d)=\emptyset$.
    \begin{proof}
        Let us consider the degree-$d$ monomial $\prod_{i=0}^{w-1}\prod_{j=1}^{l} x_{i,j}^{\alpha_{i, j}}\cdot y^{\alpha_h}$, let $d_i=\sum_{j+1}^l \alpha_{i,j}$ be its degree restricted to the $i$-th window and $D_1\ge D_2$ be the first and second maximums of $\{d_i\}$. The condition of non-admissibility requires $D_1+D_2+d_h\le 2s-1$, which implies $D_2\le \frac{D_1+D_2}{2}\le \frac{D_1+D_2=d_h}{2}\le s-1$. Therefore
        \[
        d=d_h+\sum_{i} d_i\le D_1+D_2+(w-2)D_2+d_h \le 2s-1+(w-2)(s-1)=(s-1)w+1.
        \]
        Hence, for any $d>(s-1)w+1$, $\mathcal{M}_{NA}(d)$ does not contain any monomial.
    \end{proof}
\end{proposition}
It immediately follows from the definition of the Hilbert series that
\[
\HF_{R/\mathcal{I}}(d) \ge \card{\mathcal{M}_{NA}(d)}.
\]
Experimentally, we found out that, beside non-admissible monomials, the homogenized system $S^{(y)}$ behaves as a semi-regular one. More precisely,
\begin{heuristic}
Let $S_{\mathsf{init}}=\{f_1,\dots,f_m\}\subseteq R=\F_2[x_1,\dots,x_n]$ be a randomly picked system, with $n=2^s w$. Let $S_{\mathsf{init}}'=\{\bar{g}(f_1),\dots, \bar{g}(f_m)\}\subseteq R'=\F_2[x_1',\dots,x_{sw}']$ be set 
 of polynomials obtained by applying the map $\bar{g}$ of Equation~\eqref {eq: bar_g} to $S_{\mathsf{init}}$ and $S_{\mathsf{FE}}'$ the set of polynomials corresponding to field equations for $\xv'$'s entries. Define the homogeneous ideal $\mathcal{I}=\langle (S_{\mathsf{init}}')^{(y)} \cup (S_{\mathsf{FE}}')^{(y)} \rangle$.
 Then with high-probability
 \[
 \HF_{R'/\cI}(d)=\max\left([z^d] \left[\frac{(1+z)^{sw}}{(1-z)(1+z^{2s})^{sw})}\right]_+, \card{\mathcal{M}_{NA}(d)}\right),
 \]
 where $[z^d]h(z)$ is the coefficient of the term of degree $d$ of the series $h(z)$.
\end{heuristic}

The existence of non-admissible monomials potentially increases the degree of regularity of the homogenized system. In practice, it is reasonable to consider that this does not reflect on a higher solving degree. Indeed, when solving the affine system with XL algorithm or through a Gr\"obner basis computation. On the opposite, we experimentally notice that for small concrete parameters, the degree falls may even let the solving degree decrease. Table~\ref{table: 2s rnd_vs_reg} provides a comparison of the experimental solving degree and times for some small parameters between random instances of degree-$2s$ polynomial systems and systems originated from boolean regular multivariate quadratic instances with the procedure explained above. We first remark that for random degree-$2s$ systems, the solving degree always coincides with the degree of regularity of the system \textit{before homogenizing}, i.e. with the degree of regularity of the part of highest degree, which is consistent with the possibility of exploiting degree falls during teh Gr\"obner basis computation. Moreover, we can see that the smaller solving degree is also generally reflected in a smaller time complexity. An exception is given by the case $s=1$, i.e. $l=2$, for which the results coincide. This is not surprising, for $l=2$ the solution has the same number of 0s and 1s, which means that the RMQ problem could be rewritten as the standard MQ problem with half the number of variables. A smaller complexity with our alternative approach would thus imply a better Gr\"obner basis algorithm for the multivariate quadratic problem.

\begin{table}
\begin{tabular}{ |p{1cm}||p{1cm}|p{1cm}|p{2cm}|p{1cm}|p{1cm}|p{2cm}|p{1cm}|p{1cm}|p{2cm}|}
 \hline
& \multicolumn{3}{|c|}{Quadratic modeling} & \multicolumn{3}{|c|}{Alternative modeling}  &  \multicolumn{3}{|c|}{Random degree-$(2s)$} \\ \hline
 $(s,w)$ & $d_{solv}$ & Time & Matrix &  $d_{solv}$ & Time & Matrix &  $d_{solv}$ & Time & Matrix\\ \hline \hline
$(1,20)$ & $5$ & $0.92$ & $1.1\cdot 10^5$ & $5$ & $0.33$ & $2.3\cdot 10^4$ & $5$ & $0.34$ & $2.4\cdot 10^4$ \\ \hline
$(1,25)$ & $5$ & $270$ & $3.8\cdot 10^5$ & $5$ & $9.1$ & $6.0\cdot 10^4$ & $5$ & $9.1$ & $5.8\cdot 10^4$\\ \hline
$(1,30)$ & $6$ & $7.1\cdot 10^3$ & $3.1\cdot 10^6$ & $6$ & $410$ & $6.3\cdot 10^5$ & $6$ & $420$ & $6.3\cdot 10^5$\\ \hline
$(2,5)$ & $4$ & $0.01$ & $2.0\cdot 10^3$ & $5$ & $0$ & $868$ & $6$ & $0.01$ & $640$\\ \hline
$(2,6)$ & $4$ & $0.01$ & $7.4 \cdot 10^3$ & $6$ & $0.06$ & $3.2\cdot 10^3$ & $7$ & $0.1$ & $3.3\cdot 10^3$\\ \hline
$(2,7)$ & $5$ & $0.25$ & $3.1\cdot 10^4$ & $6$ & $0.35$ & $6.1\cdot 10^3$ & $8$ & $1.23$ & $1.0 10^4$\\ \hline
$(2,8)$ & $5$ & $1.0$ & $1.0\cdot 10^5$ & $6 (7)$ & $2.1$ & $3.7\cdot 10^4$ & $8$ & $22$ & $7.1\cdot 10^4$\\ \hline
$(2,9)$ & $5$ & $4.8$ & $8.9 \cdot 10^4$ & $7$ & $40$ & $1.3\cdot 10^5$ & $9$ & $310 $& $1.5\cdot 10^5$\\ \hline
$(2,10)$ & $5$ & $63$ & $4.3\cdot 10^5$ & $7$ & $360$ & $1.2\cdot 10^5$ & $9$ & $1.0\cdot 10^4$ & $8.6\cdot 10^5$\\ \hline
$(2,11)$ & $6$ & $530$ & $1.0\cdot 10^6$ & $8$ & $5.8\cdot 10^3$ & $8.6\cdot 10^5$ & & & \\ \hline
$(2,12)$ & $6$ & $2.8\cdot 10^3$ & $1.7\cdot 10^6$ & $8$ & $5.0\cdot 10^4$ & $3.6\cdot 10^5$ & & & \\ \hline
$(2,13)$ & $6$ & $1.7\cdot 10^4$ & $2.6\cdot 10^6$ & $\ge 8$ & & $\ge 4.0\cdot 10^5$ & & & \\ \hline
$(3,4)$ & $4$ & $0.1$ & $1.9\cdot 10^4$ & $8$ & $0.21$ & $5.4\cdot 10^3$ & $9$ & $0.28$& $5.1\cdot 10^3$ \\ \hline
$(3,5)$ & $5$ & $15$ & $2.4\cdot 10^5$ & $9$ & $15$ & $4.8 \cdot 10^4$ & $10$ & $20$ & $3.0\cdot 10^4$\\ \hline
$(3,6)$ & $5$ & $170$ & $6.7\cdot 10^5$ & $9$ & $710$ & $2.4 \cdot 10^5$ & $11$ & $4.3\cdot 10^3$ & $2.2\cdot 10^5$\\ \hline
$(3,7)$ & $6$ & $3.1\cdot 10^4$ & $7.4\cdot 10^6$ & $\ge 9$ &  & $\ge 12 \cdot 10^5$ & $\ge 12$& & \\ \hline
$(4,2)$ & $3$ & $0.01$ & $6.7\cdot 10^3$ & $8$ & $0$ & $300$ & $8$ & $0.01$& $170$\\ \hline
$(4,3)$ & $4$ & $1.2$ & $9.5\cdot 10^4$ & $9$ & $0.27$ & $3.3 \cdot 10^3$ & $10$& $0.3$& $4.4\cdot 10^3$\\ \hline
$(4,4)$ & $4$ & $190$ & $3.7\cdot 10^5$ & $10$ & $86$ & $5.5 \cdot 10^4$ & $12$ & $180$& $4.6\cdot 10^4$ \\ \hline
$(5,2)$ & $3$ & $0.31$ & $7.4\cdot 10^4$ & $10$ & $0.02$ & $1.1\cdot 10^3$ & $10$ & $0.02$& $980$\\ \hline
$(5,3)$ & $4$ & $82$ & $1.5\cdot 10^6$ & $11$ & $32$ & $4.9 \cdot 10^4$ & $13$& $36$& $2.6\cdot 10^4$\\ \hline
\end{tabular}

\caption{Comparison of concrete complexities of the plain Gr\"obner basis approach for the standard quadratic modeling, the alternative degree-$(2s)$ modeling and a random degree-$(2s)$ system with $lw$ unknowns. For a comparison with the alternative modeling we only pick regular instances with $w$ blocks of length $l=2^s$. The ratio $\mu=1.2$ is slightly above the uniqueness bound to guarantee with high probability the uniqueness of the (forced) regular solution. The random system instead is forced to have a random solution. The ``$d_{solv}$'' column stands for the maximum degree reached by the MAGMA built-in Gr\"obner basis function. The time is computed in seconds and the column ``Rows'' denotes the largest number of rows of a (Macaulay) matrix computed during the Gr\"obner basis computation. The calculations have been run on one
core of an AMD EPYC 9374F processor with the MAGMA version V2.28-7.}  \label{table: 2s rnd_vs_reg}
\end{table}


In any case, we expect the degree of regularity to grow asymptotically in a very similar manner to the one for a random system of equations of degree $2s$ with $m$ equations and $sw$ unknowns, and we use it to benchmark the cost of this alternative approach. In other words, we expect the our analysis to give a pretty tight upper bound on the effective complexity of this method.

\subsubsection{Hybrid approaches}

As well as for the standard modeling, the Gr\"obner basis computation can be sped up by specializing some variables, at the cost of repeating the process until a good guess is found. The analogy goes further, as there exist a correspondence between specializing a variable in the alternative modeling and fixing several variables in the standard modeling according to the methods explained before. Indeed, let us restrict to the analysis of one window. From the definition of the bijective map $g\colon \F_2[x_1,\dots,x_l] \to \F_2[x_1',\dots,x_s']$, we have the following equivalences:
\[
\begin{array}{ccc}
    x_a'=0 & \iff & \forall j \text{ s.t. } \mathsf{bin}_a(j-1)=0,\; x_j=0,\\
    x_a'=1 & \iff & \forall j \text{ s.t. } \mathsf{bin}_a(j-1)=1,\; x_j=0,\\
\end{array}
\]
i.e. specializing one variable in the alternative modeling coincides with guessing $2^{s-1}=l/2$ zeros in a same window for the standard modeling, as the $a$-th binary digits of an element in $\Iintv{0}{2^s-1}$ is equally distributed between the values 0 and 1. Moreover, these $s$ distributions for $a\in \Iintv{1}{s}$ are mutually independent, hence specializing $s-s'$ unknowns from a same block corresponds to choosing $\sum_{i=1}^{s'} 2^{s-i}$ zero positions. In particular, guessing all $s$ variables reduces to guess all the $2^s-1$ zeros, i.e. the nonzero position. The guessing strategies are therefore very similar, however the alternative modeling admits a less refined choice for partial-window guess, since only a number of variables that is a power of 2 can left unguessed in each window.

The hybrid approach corresponding to full-window guess boils down, for the alternative modeling, to determine the optimal value $0\le \gamma \le 1$ such that specializing all $\gamma sw$ variables from $\gamma w$ blocks minimizes the complexity of solving the system. 

Therefore we study the Hilbert series

\[
\HS_{R/\mathcal{I}}(z)=\left[\frac{(1+z)^{(1-\gamma)sw}}{(1-z)(1+z^{2s})^m}\right]_{+}.
\]

We can thus write the Hilbert function at degree $d$ as
\[
\HF_{R/\mathcal{I}}(d)=I_d(n)
\]
for the Cauchy integrals 
\[
I_d(n)\eqdef \frac{1}{2\pi i}\oint \frac{1}{z^{d+1}}\frac{(1+z)^{(1-\gamma)sw}}{(1-z)(1+z^{2s})^m} dz.
\]

We set $f(z)$ such that $e^{nf(z)}=  \frac{1}{z^{d+1}}\frac{(1+z)^{(1-\gamma)sw}}{(1-z)(1+z^{2s})^m}$, i.e.
\[
f(z) \eqdef \frac{(1-\gamma)sw}{n} \ln(1+z)-\mu \ln(1+z^{2s})-\frac{1}{n}\ln(1-z)-\frac{d+1}{n}\ln(z),
\]
with $\mu =m/n$.
We compute the saddle points as
\begin{align*}
zf'(z)=&\frac{(1-\gamma)sw}{n}\frac{z}{1+z}-2s\mu \frac{z^{2s}}{1+z^{2s}}+\frac{1}{n}\frac{z}{1-z}-\frac{d+1}{n}.
\end{align*}
We observe that the term $\frac{1}{n}\frac{z}{1-z}$ is negligible when $n\to \infty$. Hence, after multiplying by all denominators and defining the degree-$(2\log(l)+1)$ polynomial
\begin{align*}
g(z)=&\frac{(1-\gamma)\log(l)w}{n}z(1+z^{2\log(l)})-2\log(l)\mu z^{2\log(l)}(1+z)-\frac{d+1}{n}(1+z)(1+z^{2\log(l)})\\
=&\left(\frac{(1-\gamma)\log(l)w}{n}-2\log(l)\mu -\frac{d+1}{n}\right)z^{2\log(l)+1}+\left(-2\log(l)\mu -\frac{d+1}{n}\right)z^{2\log(l)}\\
& +\left(\frac{(1-\gamma)\log(l)w}{n} -\frac{d+1}{n}\right)z -\frac{d+1}{n},
\end{align*}
the asymptotic relative degree of regularity $\delta\eqdef\frac{d+1}{n}$  is given by the smallest positive root of the resultant $\mathrm{Res}(g,g')$, whose degree is $2\deg(g)-1=4\log(l)+1$:
\[
\bar{\delta}= \min \{\delta \in \R \mid (\mathrm{Res}(g,g'))(\delta)=0, \quad \delta >0\}.
\]
The final asymptotic complexity thus becomes
\[
\mathcal{O}^*\left(2^{\gamma \frac{\log(l)}{l}n}\cdot \binom{\frac{(1-\gamma)\log(l)}{l}n}{\bar{\delta}n}^\omega \right)=\mathcal{O}^*\left(2^{\tau n} \right),
\]
with $\tau=\gamma \frac{\log(l)}{l}+\omega (1-\gamma)\frac{\log(l)}{l}H\left(\frac{\bar{\delta}\cdot l}{(1-\gamma) \log(l)}\right)$.

For the partial-window strategy, we notice that specializing variables decreases the degree of the terms involving that variables blocks. Since the same amount $s-s'$ of variables is guessed in each block, the degree decreases from $2s$ to $2s'$.
The relevant Hilbert series becomes instead

\[
\HS_{R/\mathcal{I}}(z)=\left[\frac{(1+z)^{s'w}}{(1-z)(1+z^{2s'})^m}\right]_{+}.
\]

We can thus write the Hilbert function at degree $d$ as
\[
\HF_{R/\mathcal{I}}(d)=I_d(n)
\]
for the Cauchy integrals 
\[
I_d(n)\eqdef \frac{1}{2\pi i}\oint \frac{1}{z^{d+1}}\frac{(1+z)^{s'w}}{(1-z)(1+z^{2s})^m} dz.
\]

We set $f(z)$ such that $e^{nf(z)}=  \frac{1}{z^{d+1}}\frac{(1+z)^{s'w}}{(1-z)(1+z^{2s'})^m}$, i.e.
\[
f(z) \eqdef \frac{s'w}{n} \ln(1+z)-\mu \ln(1+z^{2s'})-\frac{1}{n}\ln(1-z)-\frac{d+1}{n}\ln(z),
\]
with $\mu =m/n$.
We compute the saddle points as
\begin{align*}
zf'(z)=&\frac{s'w}{n}\frac{z}{1+z}-2s'\mu \frac{z^{2s'}}{1+z^{2s'}}+\frac{1}{n}\frac{z}{1-z}-\frac{d+1}{n}.
\end{align*}
We observe that the term $\frac{1}{n}\frac{z}{1-z}$ is negligible when $n\to \infty$. Let $l'=s^{s'}$. After multiplying by all denominators and defining the degree-$(2\log(l')+1)$ polynomial
\begin{align*}
g(z)=&\frac{\log(l')w}{n}z(1+z^{2\log(l')})-2\log(l')\mu z^{2\log(l')}(1+z)-\frac{d+1}{n}(1+z)(1+z^{2\log(l')})\\
=&\left(\frac{\log(l')w}{n}-2\log(l')\mu -\frac{d+1}{n}\right)z^{2\log(l')+1}+\left(-2\log(l')\mu -\frac{d+1}{n}\right)z^{2\log(l')}\\
& +\left(\frac{\log(l')w}{n} -\frac{d+1}{n}\right)z -\frac{d+1}{n},
\end{align*}
the asymptotic relative degree of regularity $\delta\eqdef\frac{d+1}{n}$  is given by the smallest positive root of the resultant $\mathrm{Res}(g,g')$, whose degree is $2\deg(g)-1=4\log(l')+1$:
\[
\bar{\delta}= \min \{\delta \in \R \mid (\mathrm{Res}(g,g'))(\delta)=0, \quad \delta >0\}.
\]
The final asymptotic complexity thus becomes
\[
\mathcal{O}^*\left(2^{\frac{\log(l/l')}{l}n}\cdot \binom{\frac{\log(l')}{l}n}{\bar{\delta}n}^\omega \right)=\mathcal{O}^*\left(2^{\tau n} \right),
\]
with $\tau= \frac{\log(l')}{l}+\omega \frac{\log(l')}{l}H\left(\frac{\bar{\delta}\cdot l}{ \log(l')}\right)$.

\subsection{Polynomial method on the alternative modeling}
In our previous analysis of the probabilistic polynomial method, we focused on the case of quadratic systems. The general idea of the algorithm, however, works for equations of higher degree as well and each of the already mentioned works \cite{LPTWY17, BKW19, D20, D21} provides the estimated complexity in function of the degree. The aim of this subsection is to provide the complexity of the polynomial method to the system of $m$ equations of degree $2s$ in $sw$ variables turning out from the application of the map $\bar{g}$. Since the solutions are no longer subject to the regular constraints, the cost estimate is a mere application of the best result among all the probabilistic method variants with respect to polynomials of degree $2s$. In particular the recursive version of \cite{D20} holds the asymptotic complexity record of $\mathcal{O}^*(2^{(1-1/(2d))n})$ for any degree $d>2$. We recall that \cite{D20} was also the best for degree $2$. When applied to our system, we thus obtain a complexity of
\[
\mathcal{O}^*(2^{(1-1/(4s))sw})=\mathcal{O}^*\left(2^{\left(\frac{\log(l)}{l}-\frac{1}{4l}\right)n}\right).
\]

\section{Comparison}
We provide here below figures with the asymptotic complexities for the plain and hybrid Gr\"obner basis approaches assuming $\omega=2$ and $\mu=\log(l)/l$, as well as for the polynomial methods and we provide the exhaustive search as a benchmark. We show both the absolute complexity and the relative complexity with respect to brute force, i.e. where we reparametrize in such a way that brute force has unitary cost. In the latter case, we note that plain Gr\"obner basis does not tend to 1. The following explanation provides an intuition of this behavior. The extremal case is when there is only one window and the total weight is 1. Indeed, the absolute degree of regularity tends to 2 and thus the (nonconstant) complexity coefficient is given by
$\omega H(2/n)$, while for brute force it is clearly $\log(n)$. We thus obtain
\[
\lim_{n\to \infty} \frac{\omega H(2/n)}{\log(n)}=\lim_{n\to \infty} \frac{\omega (2\log(n)-2-(n-2)\log(n)+(n-2)\log(n-2))}{\log(n)}=2\omega=4.
\]
Note that for $l=2$ the different window guess coincides with the full window guess, because either 0 or 1 entries per window are guessed. For the same reason, the hybrid partial-window guess is nothing but the best between a plain Gr\"obner basis and the exhaustive search, with the former being better for $l=2$. However, for $l\ge 3$, the hybrid partial-window guess becomes better than the full-window guess approach.
Overall, the best asymptotic approach is the hybrid approach with a different number of guesses per window, as it is the one that makes use of more parameters to optimize the cost among the Gr\"obner basis methods. The drawback is that even finding the best parameters may become non-trivial as $l$ increases. On the other hand, we observe that already for $l\ge 4$, the method seems to coincide (or almost coincide) with the sub-case of hybrid partial-window guess, and we can thus consider the latter to quite reliably estimate the asymptotic complexity for a fixed $l$.

We remark that the probabilistic methods can still be relevant in regimes other than the one analyzed here for $\mu=\frac{\log(l)}{l}$. Indeed, differently from a Gr\"obner basis computation, whose complexity highly varies with the number of equations, the formers are mostly independent from them. Therefore, if the system is enough underdetermined, then the probabilistic methods may still outperform asymptotically Gr\"obner basis approaches in finding \text{a} regular solution. 

\begin{figure}
    \centering
    \begin{tikzpicture}[scale=0.70]
  \begin{axis}[
    xlabel=window length $l$,
    ylabel=coefficient $\tau$ in complexity $2^{\tau n}$, xmode=log, legend style={cells={align=center}}]
    \addplot coordinates {(2, 0.4364) (3, 0.5484) (4, 0.5773) (5, 0.5773) (6, 0.5659) (10, 0.4982) (20, 0.3710) (50, 0.2179) (100, 0.1355) };

            \addplot coordinates {(2, 0.5) (3, 0.5283) (4, 0.5) (5, 0.4644) (6, 0.4308) (10, 0.3322) (20, 0.2161) (50, 0.1129) (100, 0.06644) };
        \addplot coordinates {(2, 0.3955) (3, 0.4567) (4, 0.4512) (5, 0.4297) (6, 0.4052) (10, 0.3220) (20, 0.2136) (50, 0.1125) (100, 0.06637)};
        \addplot coordinates {(2, 0.4364) (3, 0.4195) (4, 0.3962) (5, 0.3708) (6, 0.3469) (10, 0.2741) (20, 0.1836) (50, 0.09826) (100, 0.05831)};
        \addplot coordinates {(2, 0.3955) (3, 0.4179) (4, 0.3962) (5, 0.3708) (6, 0.3469)};
        \addplot coordinates {(2, 0.4272) (3, 0.4798) (4, 0.4427) (5, 0.4185) (6, 0.3852) (10, 0.3002) (20, 0.1979) (50, 0.1048) (100, 0.06221)}; 
        \addplot coordinates {(2, 0.4249) (3, 0.4783) (4, 0.4415) (5, 0.4175) (6, 0.3843) (10, 0.2997) (20, 0.1976) (50, 0.1047) (100, 0.06219)}; 
        \addplot coordinates {(2, 0.4075) (3, 0.4667) (4, 0.4286) (5, 0.4073) (6, 0.3741) (10, 0.2927) (20, 0.1940) (50, 0.1034) (100, 0.06156)}; 
   \legend{plain GB, brute force, hybrid full-window guess, hybrid partial-window guess, hybrid different windows, polynomial method (no recursion), polynomial method \\(recursion \`a la Bjorklund et al.), polynomial method \\(recursion \`a la Dinur)}
  \end{axis}
\end{tikzpicture}
    \caption{Absolute complexity, comparison among algebraic and probabilistic methods}
    \label{fig:abs_alt}
\end{figure}
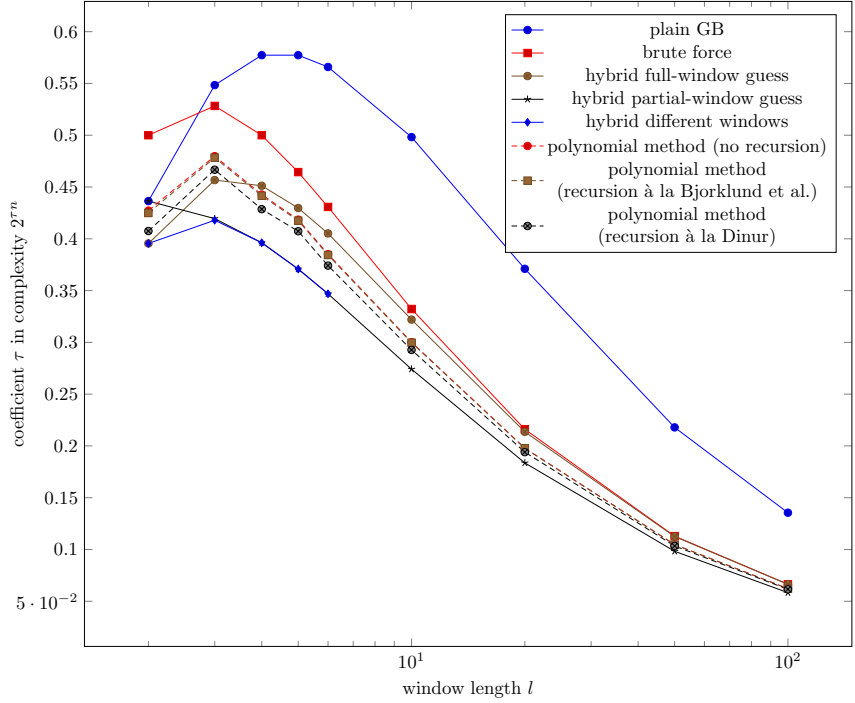

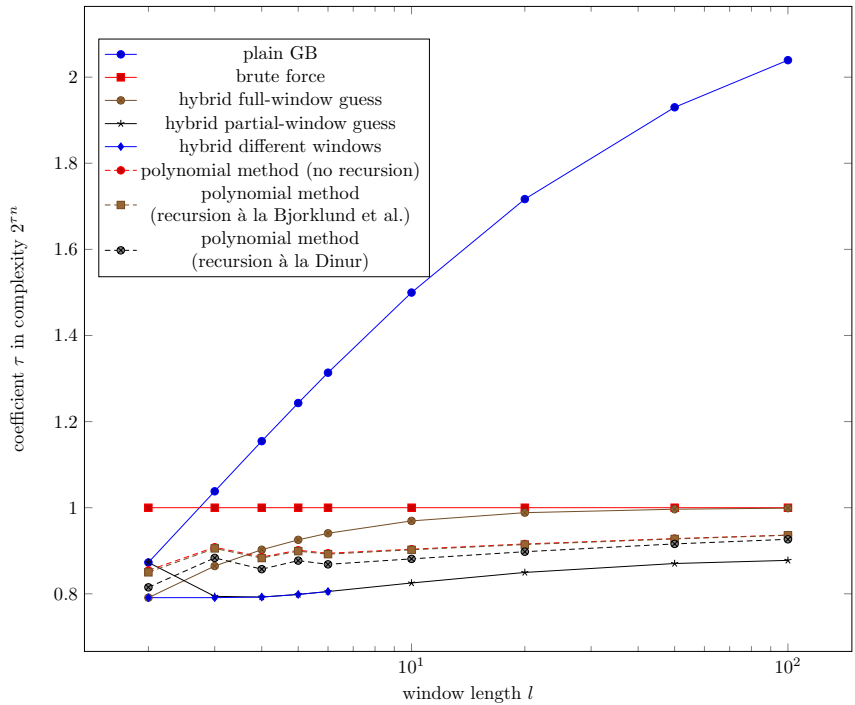
\begin{figure}
    \centering
    \begin{tikzpicture}[scale=0.70]
  \begin{axis}[
    xlabel=window length $l$,
    ylabel=coefficient $\tau$ in complexity $2^{\tau n}$, xmode=log,  legend style={cells={align=center}, at={(0.45,0.95)}}]
    \addplot coordinates {(2, 0.4364/0.5) (3, 0.5484/0.5283) (4, 0.5773/0.5) (5, 0.5773/0.4644) (6, 0.5659/0.4308) (10, 0.4982/0.3322) (20, 0.3710/0.2161) (50, 0.2179/0.1129) (100, 0.1355/0.06644) };

            \addplot coordinates {(2, 0.5/0.5) (3, 0.5283/0.5283) (4, 0.5/0.5) (5, 0.4644/0.4644) (6, 0.4308/0.4308) (10, 0.3322/0.3322) (20, 0.2161/0.2161) (50, 0.1129/0.1129) (100, 0.06644/0.06644) };
        \addplot coordinates {(2, 0.3955/0.5) (3, 0.4567/0.5283) (4, 0.4512/0.5) (5, 0.4297/0.4644) (6, 0.4052/0.4308) (10, 0.3220/0.3322) (20, 0.2136/0.2161) (50, 0.1125/0.1129) (100, 0.06637/0.06644)};
        \addplot coordinates {(2, 0.4364/0.5) (3, 0.4195/0.5283) (4, 0.3962/0.5) (5, 0.3708/0.4644) (6, 0.3469/0.4308) (10, 0.2741/0.3322) (20, 0.1836/0.2161) (50, 0.09826/0.1129) (100, 0.05831/0.06644)};
        \addplot coordinates {(2, 0.3955/0.5) (3, 0.4179/0.5283) (4, 0.3962/0.5) (5, 0.3708/0.4644) (6, 0.3469/0.4308)};
        \addplot coordinates {(2, 0.4272/0.5) (3, 0.4798/0.5283) (4, 0.4427/0.5) (5, 0.4185/0.4644) (6, 0.3852/0.4308) (10, 0.3002/0.3322) (20, 0.1979/0.2161) (50, 0.1048/0.1129) (100, 0.06221/0.06644)}; 
        \addplot coordinates {(2, 0.4249/0.5) (3, 0.4783/0.5283) (4, 0.4415/0.5) (5, 0.4175/0.4644) (6, 0.3843/0.4308) (10, 0.2997/0.3322) (20, 0.1976/0.2161) (50, 0.1047/0.1129) (100, 0.06219/0.06644)}; 
        \addplot coordinates {(2, 0.4075/0.5) (3, 0.4667/0.5283) (4, 0.4286/0.5) (5, 0.4073/0.4644) (6, 0.3741/0.4308) (10, 0.2927/0.3322) (20, 0.1940/0.2161) (50, 0.1034/0.1129) (100, 0.06156/0.06644)}; 
   \legend{plain GB, brute force, hybrid full-window guess, hybrid partial-window guess, hybrid different windows, polynomial method (no recursion), polynomial method \\(recursion \`a la Bjorklund et al.), polynomial method \\(recursion \`a la Dinur)}
  \end{axis}
\end{tikzpicture}
    \caption{Complexity relative to brute force, comparison among algebraic and probabilistic methods}
    \label{fig:abs_alt}
\end{figure}

\begin{figure}
    \centering
    \begin{tikzpicture}[scale=0.70]
  \begin{axis}[
    title=complexity relative to brute force,
    xlabel=window length $l$,
    ylabel=coefficient $\tau$ in complexity $2^{\tau n}$, xmode=log,  legend style={cells={align=center}, at={(0.95,0.73)}}]
    \addplot coordinates {(2, 0.4364/0.5) (4, 0.5773/0.5) (8, 0.5328/0.3750) (16, 0.4130/0.25) (32, 0.2870/0.1563) (64, 0.1851/0.09375) (128, 0.1132/0.05469) };
        \addplot coordinates {(2, 0.3955/0.5)  (4, 0.3962/0.5) (8, 0.3062/0.3750) (16, 0.2105/0.25) (32, 0.1348/0.1563) (64, 0.08184/0.09375) (128, 0.04814/0.05469) };
        \addplot coordinates {(2, 0.4364/0.5)  (4, 0.7244/0.5) (8, 0.6235/0.3750) (16, 0.4421/0.25) (32, 0.2858/0.1563) (64, 0.1751/0.09375) (128, 0.1036/0.05469) };
        \addplot coordinates {(2, 0.3955/0.5)  (4, 0.4970/0.5) (8, 0.3750/0.3750) (16, 0.2500/0.25) (32, 0.1562/0.1563) (64, 0.09375/0.09375) (128, 0.05469/0.05469) };
        \addplot coordinates {(2, 0.4364/0.5)  (4, 0.3962/0.5) (8, 0.3062/0.3750) (16, 0.2105/0.25) (32, 0.1348/0.1563) (64, 0.08242/0.09375) (128, 0.04879/0.05469) };
        \addplot coordinates {(2, 0.4075/0.5) (4, 0.4286/0.5) (8, 0.3284/0.3750) (16, 0.2231/0.25) (32, 0.1418/0.1563) (64, 0.08625/0.09375) (128, 0.05086/0.05469)}; 
        \addplot coordinates {(2, 0.3750/0.5) (4, 0.4375/0.5) (8, 0.3438/0.3750) (16, 0.2344/0.25) (32, 0.1484/0.1563) (64, 0.08984/0.09375) (128, 0.05273/0.05469)}; 
   \legend{plain GB, hybrid GB (full/partial), plain GB \\(alternative modeling), hybrid full-window\\(alternative modeling), hybrid partial-window\\(alternative modeling), polynomial method \\(recursion \`a la Dinur),  polynomial method \\(recursion \`a la Dinur\\alternative modeling)}
  \end{axis}
\end{tikzpicture}
    \caption{Complexity relative to brute force, comparison with the methods for the alternative system description}
    \label{fig:rel_alt}
\end{figure}
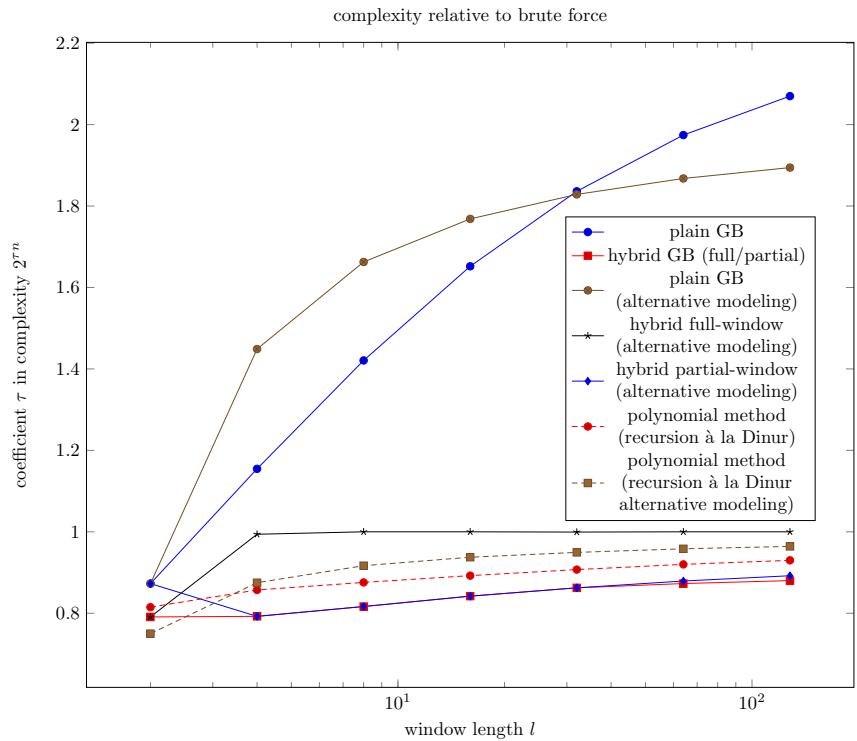

Regarding the complexity for the solvers of the alternative modeling (Figures~\ref{fig:abs_alt} and \ref{fig:rel_alt}, we see that a plain Gr\"obner basis beats that of original modeling for big enough windows (the smallest power of 2 for which this is realized is $l=2^5$). The best value of $\gamma$ for the hybrid approach quickly tends to 1 with $l$ growing, making this method coincide with brute force and the two 2 curves almost overlap already starting from $l=8$. This is due to the growth of the polynomial degrees, the algebraic part becomes less competitive as the latter increase. For $l=\{4,8,16,32\}$, the two modelings with hybrid partial-window specialization coincide. This follows from Remark~\ref{remark: l'=2} and the fact that the RMQ problem with $l'=2$ reduces to the standard MQ problem with half the number of variables. We also note that, when $s'=1$, the (specialized) \textit{alternative} modeling has just a random structure, because of degree $2s'=2$, i.e. it can be treated as an overdetermined random quadratic system.

Due to all what have been said so far, we conclude that a pretty simple and elegant, yet accurate, estimation of the complexity of an RMQ instance at the uniqueness bound is obtained by guessing $l-2$ in each block and thus reducing the problem to many MQ instances in $w$ unknowns and $\log(l)w$ quadratic equations, i.e. with $\mu=\log(l)$. For $0.55\mu\ge 1$, which corresponds to $\l\ge 4$, we have already said that the best trade-off consists of a plain Gr\"obner basis computation. The simplicity of this analysis stems from the existence of a closed form asymptotic expression for the relative degree of regularity $\delta$. We recall indeed that \cite[Corollary 4.4.3]{B04} determines the asymptotic of $\delta$ for a semi-regular quadratic system over $\F_2$ with ratio $\mu$ as
\begin{equation} \label{eq: delta_Cor443}
\delta \to \bar{\delta}\eqdef -\mu+\frac{1}{2}+\frac{1}{2}\sqrt{2\mu^2-10\mu-1+2(\mu+2)\sqrt{\mu(\mu+2)}}.
\end{equation}
Hence an asymptotic upper bound on the complexity of solving a boolean RMQ instance with $w$ blocks of length $l$ is
\[
\mathcal{O}^*\left(\left(\frac{l}{l'}\right)^w\cdot \binom{w}{\bar{\delta} w}^\omega\right)=\mathcal{O}^*\left(2^{\left(\frac{\log(l/l')+\omega H(\bar{\delta})}{l}\right)n}\right),
\]
where $\bar{\delta}$ is defined as in \eqref{eq: delta_Cor443} with $\mu=\log(l)$.

Finally, our estimates show that the partial/different-window hybrid technique remains asymptotically the most efficient for higher values of $l$, even compared to other methods. Indeed, for $l=2$ the recursion \`a la Dinur in the probabilistic method for the alternative modeling beats any other approach, which  is consistent with the probabilistic method being asymptotically better than Gr\"obner basis techniques for MQ instances at the uniqueness bound. However, the former, when applied to either the standard or the alternative modeling, becomes more costly already for $l\ge 4$.

\begin{figure}
    \centering
    \begin{tikzpicture}[scale=0.70]
  \begin{axis}[
    title=absolute complexity,
    xlabel=window length $l$,
    ylabel=coefficient $\tau$ in complexity $2^{\tau n}$, xmode=log, legend style={cells={align=center}}]
    \addplot coordinates {(2, 0.4364) (4, 0.5773) (8, 0.5328) (16, 0.4130) (32, 0.2870) (64, 0.1851) (128, 0.1132) };
            \addplot coordinates {(2, 0.5)  (4, 0.5) (8, 0.3750) (16, 0.2500) (32, 0.1563) (64, 0.09375) (128, 0.05469) };
        \addplot coordinates {(2, 0.3955)  (4, 0.3962) (8, 0.3062) (16, 0.2105) (32, 0.1348) (64, 0.08184) (128, 0.04814) };
        \addplot coordinates {(2, 0.4364)  (4, 0.7244) (8, 0.6235) (16, 0.4421) (32, 0.2858) (64, 0.1751) (128, 0.1036) };
        \addplot coordinates {(2, 0.3955)  (4, 0.4970) (8, 0.3750) (16, 0.2500) (32, 0.1562) (64, 0.09375) (128, 0.05469) };
        \addplot coordinates {(2, 0.4364)  (4, 0.3962) (8, 0.3062) (16, 0.2105) (32, 0.1348) (64, 0.08242) (128, 0.04879) };
        \addplot coordinates {(2, 0.4075) (4, 0.4286) (8, 0.3284) (16, 0.2231) (32, 0.1418) (64, 0.08625) (128, 0.05086)}; 
        \addplot coordinates {(2, 0.3750) (4, 0.4375) (8, 0.3438) (16, 0.2344) (32, 0.1484) (64, 0.08984) (128, 0.05273)}; 
   \legend{plain GB, brute force, hybrid GB (full/partial), plain GB \\(alternative modeling),  hybrid full-window\\(alternative modeling), hybrid partial-window\\(alternative modeling),  polynomial method \\(recursion \`a la Dinur),  polynomial method \\(recursion \`a la Dinur\\alternative modeling)}
  \end{axis}
\end{tikzpicture}
    \caption{Absolute complexity, comparison with the methods for the alternative system description}
    \label{fig:abs_alt}
\end{figure}
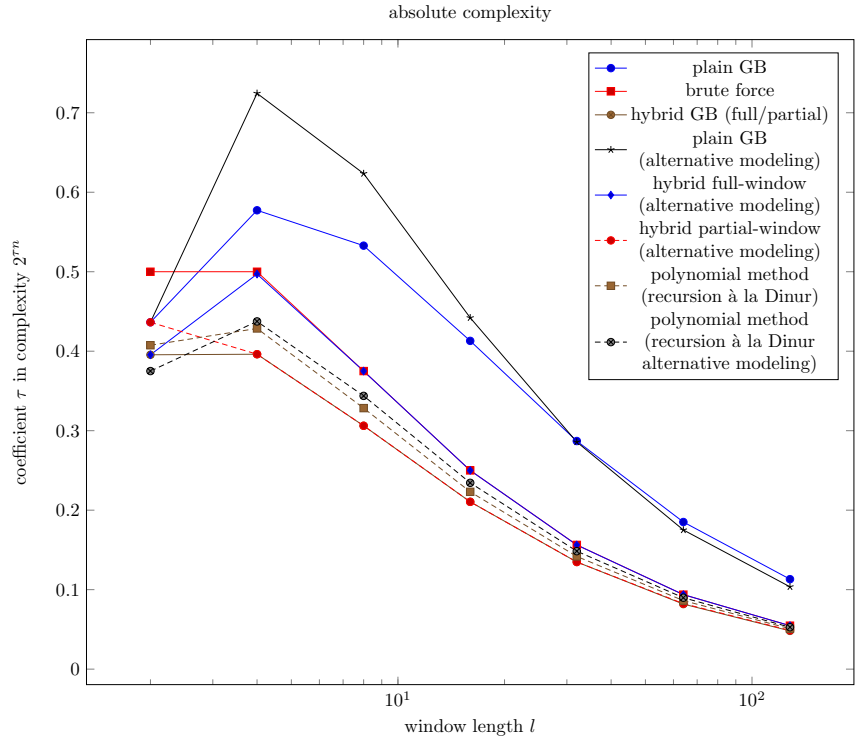

\section{Conclusions}

We presented a new NP-complete variant of the multivariate quadratic problem where, instead of considering structured polynomials, the additional property of having weight-1 blocks is required from the solution. We reviewed all the most competitive techniques for solving polynomial systems over finite fields. By doing this, we devised tailored approaches that take into account the regular constraint. Despite the non-negligible improvements over the standard multivariate quadratic problem, our analysis shows that, even in the average case, the problem remains hard. Indeed, the complexity is still exponential and, for block lengths of cryptographic interest, has a not-so-small constant at the exponent.

Algebraic cryptanalysis resulted in being, asymptotically, the most efficient method. This is surprising to some extent because it is not what happens in the usual MQ setting, where the best polynomial methods beat the hybrid Gr\"obner basis computation for sufficiently large parameters, at least in theory.
 Moreover, as remarked in the corresponding section, the best hybrid strategy is very close in essence to an exhaustive search rather than a plain Gr\"obner basis computation, meaning that the best trade-off is achieved by specializing almost all variables in each block. This fact is an additional piece of evidence that the problem is difficult in practice.

Even more so for concrete instances, we expect hybrid algebraic cryptanalysis to perform better than polynomial methods, as the latter have not found much use in practice for cryptographic parameters, not even for the MQ problem. 
In this regard, an interesting open problem would be to study the efficiency of the Crossbred algorithm with respect to RMQ and possibly adapt it to this scenario, as this is the only work that defeats fast exhaustive search for practical parameters.

The RMQ problem naturally inherits properties from MQ and RSD problems, and both revealed to be well suited for cryptographic constructions, for instance for designing digital signatures. It is then tantalizing to conjecture that the RMQ problem could have many cryptographic applications as well, and we leave for future work devising competitive protocols based on the corresponding hardness assumption.

\bibliographystyle{alpha}
\bibliography{ref}

\appendix
\section{Plain Gr\"obner basis analysis over $\F_2$} \label{app: plainF2}

Let us define the Cauchy integrals 
\[
I_d(n)\eqdef \frac{1}{2\pi i}\oint \frac{1}{z^{d+1}}\frac{(1+(l-1)z)^w}{(1-z)(1+z^2)^m} dz
\]
for all non-negative integers $d$.
Each coefficient of the Hilbert series can be written as the corresponding integral
\[
\HF_{R/\mathcal{I}}(d)=I_d(n).
\]
We set $f(z)$ such that $e^{nf(z)}= \frac{1}{z^{d+1}}\frac{(1+(l-1)z)^w}{(1-z)(1+z^2)^m}$, i.e.
\[
f(z) \eqdef \frac{w}{n} \ln(1+(l-1)z)-\frac{m}{n}\ln(1+z^2)-\frac{1}{n}\ln(1-z)-\frac{d+1}{n}\ln(z).
\]
We compute the saddle points as
\begin{align*}
zf'(z)=&\frac{(l-1)w}{n}\frac{z}{1+(l-1)z}-\frac{2m}{n}\frac{z^2}{1+z^2}+\frac{1}{n}\frac{z}{1-z}-\frac{d+1}{n}\\
=&\frac{(l-1)}{l}\frac{z}{1+(l-1)z}-2\mu\frac{z^2}{1+z^2}+\frac{1}{n}\frac{z}{1-z}-\frac{d+1}{n}.
\end{align*}
We observe that the term $\frac{1}{n}\frac{z}{1-z}$ is negligible when $n\to \infty$. Hence, the computation leads to the condition
\[
\frac{(l-1)}{l} z(1+z^2)-2\mu z^2(1+(l-1)z)-\frac{d+1}{n}(1+z^2)(1+(l-1)z)=0
\]
and we want to find the value $\delta\eqdef\frac{d+1}{n}$ which provides the asymptotic of the relative degree of regularity. The cubic equation in $z$ above can be rewritten in the canonical form
\[
P(z)=p_3 z^3+p_2 z^2+p_1 z+p_0=0
\]
where
\[
\begin{cases}
    p_3 &= (l-1)\left(2\mu +\delta-\frac{1}{l}\right) \\
    p_2 &= 2\mu +\delta\\
    p_1 &= (l-1)\left(\delta -\frac{1}{l}\right) \\
    p_0 &= \delta.
\end{cases}
\]

We want to find coalescent saddle points, i.e. a root of multiplicity 2 for $P$. For cubic polynomials, this can be found by forcing the discriminant to vanish. The discriminant is defined as
\[
\Delta(P)=18 p_3 p_2 p_1 p_0-4 p_2^3 p_0 + p_2^2 p_1^2- 4 p_3 p_1^3 -27 p_3^2 p_0^2.
\]
Equating the discriminant to 0 (and dividing by -4) provides a quartic equation
\begin{equation} \label{eq: quarticF2}
r_4 \delta^4+ r_3 \delta^3+r_2\delta^2+r_1\delta +r_0=0
\end{equation}
in the variable $\delta$, whose coefficients are polynomial functions of $\mu$ and $l$:
\[
\begin{cases}
    r_4 &= ((l-1)^2+1)^2 \\
    r_3 &= 2\mu ((l-1)^2+1)((l-1)^2+3)-4\frac{(l-1)^2((l-1)^2+1)}{l}\\
    r_2 &= 4\mu^2 (2(l-1)^2+3)- 2\mu\frac{(l-1)^2(3(l-1)^2-1)}{l} +2\frac{(l-1)^2 (3 (l-1)^2 +1)}{l^2}\\
    r_1 &= 8 \mu^3+20\mu^2 \frac{(l-1)^2}{l} + 2\mu (l-1)^2 \frac{3(l-1)^2-5}{l^2 }-4 \frac{(l-1)^4}{l^3} \\
    r_0 &= -\mu^2 \frac{(l-1)^2}{l^2}-2\mu \frac{(l-1)^4}{l^3}+\frac{(l-1)^4}{l^4}.
\end{cases}
\]
The smallest positive root $\bar{\delta}$ of \eqref{eq: quarticF2} gives the good asymptotic for the relative degree of regularity.

\section{Plain Gr\"obner basis approach over $\F_q$} \label{app: plainFq}

We can write the Hilbert function at degree $d$ as
\[
\HF_{R/\mathcal{I}}(d)=I_d(n)
\]
for the Cauchy integrals 
\[
I_d(n)\eqdef \frac{1}{2\pi i}\oint \frac{1}{z^{d+1}}\frac{(1+(l-1)z)^w (1-z^2)^m}{(1-z)^{w+1}} dz.
\]

We set $f(z)$ such that $e^{nf(z)}=  \frac{1}{z^{d+1}}\frac{(1+(l-1)z)^w (1-z^2)^m}{(1-z)^{w+1}} $, i.e.
\[
f(z) \eqdef \frac{w}{n} \ln(1+(l-1)z)-\frac{m}{n}\ln(1-z^2)-\frac{w+1}{n}\ln(1-z)-\frac{d+1}{n}\ln(z).
\]
We compute the saddle points as
\begin{align*}
zf'(z)=&\frac{(l-1)w}{n}\frac{z}{1+(l-1)z}-\frac{2m}{n}\frac{z^2}{1-z^2}+\frac{w+1}{n}\frac{z}{1-z}-\frac{d+1}{n}\\
=&\frac{(l-1)}{l}\frac{z}{1+(l-1)z}-2\mu\frac{z^2}{1-z^2}+\frac{1}{l}\frac{z}{1-z}+\frac{1}{n}\frac{z}{1-z}-\frac{d+1}{n}.
\end{align*}
We observe that the term $\frac{1}{n}\frac{z}{1-z}$ is negligible when $n\to \infty$. Hence, the computation leads to the condition
\[
\frac{(l-1)}{l} z(1-z^2)-2\mu z^2(1+(l-1)z)+\frac{1}{l}z(1+z)(1+(l-1)z)-\frac{d+1}{n}(1-z^2)(1+(l-1)z)=0
\]
and we want to find the value $\delta\eqdef\frac{d+1}{n}$ which provides the asymptotic of the relative degree of regularity. The cubic equation in $z$ above can be rewritten in the canonical form
\[
P(z)=p_3 z^3+p_2 z^2+p_1 z+p_0=0
\]
where
\[
\begin{cases}
    p_3 &= (l-1)\left(2\mu -\delta\right) \\
    p_2 &= 2\mu -\delta-1\\
    p_1 &= \delta(l-1)-1 \\
    p_0 &= \delta.
\end{cases}
\]

We want to find coalescent saddle points, i.e. a root of multiplicity 2 for $P$. For cubic polynomials, this can be found by forcing the discriminant to vanish. The discriminant is defined as
\[
\Delta(P)=18 p_3 p_2 p_1 p_0-4 p_2^3 p_0 + p_2^2 p_1^2- 4 p_3 p_1^3 -27 p_3^2 p_0^2.
\]
Equating the discriminant to 0 provides a quartic equation
\begin{equation} \label{eq: quarticFq}
r_4 \delta^4+ r_3 \delta^3+r_2\delta^2+r_1\delta +r_0=0
\end{equation}
in the variable $\delta$, whose coefficients are polynomial functions of $\mu$ and $l$:
\[
\begin{cases}
    r_4 &= 4l^2(2l+1)^2 \\
    r_3 &= -8\mu (l^3-4l^2-2l+4)-4(3l^2-8l+8)(l-2)\\
    r_2 &= -16\mu^2 (2l^2-4l-1)+ 8\mu(3l^3-14l^2+29l-24)+13l^2-48l+48\\
    r_1 &= -32\mu^3-16\mu^2 (5l-8) -4\mu (6l^2-23+24) -6(l-2) \\
    r_0 &= 4\mu^2 +4\mu(2l-3)+1.
\end{cases}
\]
The smallest positive root $\bar{\delta}$ of \eqref{eq: quarticFq} gives the good asymptotic for the relative degree of regularity.

\end{document}